\documentclass[format=acmsmall, acmtiis, manuscript,screen]{acmart}

\AtBeginDocument{%
  \providecommand\BibTeX{{%
    \normalfont B\kern-0.5em{\scshape i\kern-0.25em b}\kern-0.8em\TeX}}}

\setcopyright{acmlicensed}
\copyrightyear{2024}
\acmYear{2024}
\acmDOI{XXXXXXX.XXXXXXX}

\acmConference[Conference acronym 'XX]{Make sure to enter the correct
  conference title from your rights confirmation emai}{June 03--05,
  2024}{Woodstock, NY}
\acmISBN{978-1-4503-XXXX-X/18/06}

\begin{document}

\title[AI Drawing Partner]{AI Drawing Partner: Co-Creative Drawing Agent and Research Platform to Model Co-Creation}



\author{Nicholas Davis}
\email{nicholas.davis10@gmail.com}
\orcid{0000-0002-4427-0286}
\affiliation{
  \institution{Co-Creative AI Consulting}
  \city{Elyria}
  \state{Ohio}
  \country{USA}
  \postcode{44035}
  }

\author{Janet Rafner}
\email{janetrafner@mgmt.au.dk}
\orcid{0000-0001-9264-3334}
\affiliation{
  \institution{Center for Hybrid Intelligence, Department of Management, Aarhus University}
  \country{Denmark}
  }



\renewcommand{\shortauthors}{Short authors}

\begin{abstract}
This paper describes the AI Drawing Partner, which is a co-creative drawing agent that also serves as a research platform to model co-creation. The AI Drawing Partner is an early example of a quantified co-creative AI system that automatically models the co-creation that happens on the system. The method the system uses to capture this data is based on a new cognitive science framework called co-creative sense-making (CCSM). The CCSM is based on the cognitive theory of enaction, which describes how meaning emerges through interaction with the environment and other people in that environment in a process of sense-making. The CCSM quantifies elements of interaction dynamics to identify sense-making patterns and interaction trends. This paper describes a new technique for modeling the interaction and collaboration dynamics of co-creative AI systems with the co-creative sense-making (CCSM) framework. There are four categories of data collected in the CCSM: cognitive dynamics, interaction dynamics, collaboration dynamics, and domain behaviors. The AI Drawing Partner uses these categories as a data collection schema for quantifying co-creation. The system’s interaction design and software architecture are presented. A case study is conducted of ten co-creative drawing sessions between a human user and the co-creative agent. Five of the sessions are abstract and five are representational. The analysis includes showing the artworks produced, the quantified data from the AI Drawing Partner, the curves describing interaction dynamics, and a visualization of interaction trend sequences. The results help validate the CCSM by showing significant differences in the interaction dynamics and collaboration dynamics between abstract and representational sessions that match the qualitative difference of the user’s interactions with the system. The primary contribution of this paper is presenting the AI Drawing Partner, which is a unique co-creative AI system and research platform that collaborates with the user in addition to quantifying, modeling, and visualizing the co-creative process using the CCSM framework.

\end{abstract}



\received{1 May 2024}

\maketitle

\section{Introduction}

Generative AI (genAI) has the potential to radically transform how we engage with technology. The computational power of genAI paired with the creativity of the user can accelerate the user's creative process \cite{bilgram2023accelerating}. Yet, to ensure control, collaboration, and the construction of shared meanings, these systems require a design that centers on human needs and perspectives \cite{chen2023next, rafner2023creativity, mao2023hybrid}. Co-creative artificial intelligence (AI) is a field of study that combines computational creativity \cite{colton2008creativity,colton2011computational,colton2012computational,jordanous2016four,wiggins2006searching},  human-computer interaction (HCI) \cite{suchman2007human,hutchins2000distributed,hollan2000distributed,carroll1997human,norman2013design}, and insights from psychological sciences \cite{rafner2023creativity} to investigate and design systems where the user and agent collaborate on a shared creative product \cite{davis2013human}. Recent publicly accessible examples of generative AI, such as ChatGPT \cite{chatGPT}, Github Copilot \cite{copilot}, and Stable Diffusion \cite{rombach2022high}, are examples of co-creative systems. Co-creative systems are a subset of mixed-initiative creative interfaces (MICI) that share the creative initiative between the user and system \cite{deterding2017mixed,yannakakis2014mixed}. There are co-creative systems in a wide variety of creative domains, such as drawing \cite{davis2013human, fan2019collabdraw,deshpande2021drawcto}, design \cite{karimi2019deep,ibarrola2023collaborative, zhu2018explainable,kim2021evaluating}, poetry \cite{kantosalo2019usability,oliveira2019co,hamalainen2018poem}, music \cite{biles2002genjam,huang2020ai,louie2020novice}, theatrical improvisation \cite{magerko2011digital}, and game design \cite{guzdial2018co,margarido2022let}. 

In co-creative AI, both simultaneous collaboration and turn-taking can emerge between the user and the co-creative agent \cite{winston2017turn}. Turns can be mutually influential and lead to a structural coupling between the user and co-creative agent, where there is some structural correspondence between a series of turns (e.g., building up the components of a face in a drawing would be a structural coupling) \cite{davis2016empirically}. In co-creative experiences, the agent and user can engage in an improvisational collaboration, where contributions can build upon what their partner contributed \cite{davis2013human}. In this context, meaning is dynamic and grows through interaction \cite{davis2015enactive}. 

This paper describes the AI Drawing Partner, which is a co-creative drawing agent that collaborates with the user in real time on a shared canvas. The agent selects its actions based on analyzing the user’s line input and preferences via voting over time. The AI Drawing Partner explores how to situate genAI algorithms in a human-centered co-creative drawing context. Current image generation systems, such as DALLE-2 \cite{dalle2}, Google's Gemini \cite{gemini}, and Imagine with Meta AI \cite{meta}, generate one or several static images that the user can download. However, the user cannot edit, manipulate, re-size, or sketch over the image, limiting human control, a key point of discussion in human-centered AI frameworks \cite{shneiderman2020human}. The AI Drawing Partner enables image generation in the context of creating a shared artwork, meaning the images can be manipulated and edited on the canvas. The AI Drawing Partner also features communication between the user and agent to build trust and enable control, such as positive/negative feedback, requesting sketches/images, and changing drawing modes. 

The AI Drawing Partner has built in recording metrics that automatically capture interaction details specified in the co-creative sense-making cognitive framework (CCSM), effectively coding the interaction. The CCSM combines frameworks from cognitive science to provide a new cognitive framework and data collection schema for co-creative systems to implement. These quantified co-creative systems automatically code the interactions of collaborators and quantify interaction, collaboration, cognitive, and domain  dynamics. This approach significantly reduces the amount of time it takes to analyze the data and improves data reliability by removing human bias and error from the coding process. The AI Drawing Partner can serve as an experimental platform for collecting data about the co-creative visual artistic process. The system’s data collection mechanisms can serve as a model for how to quantify co-creation in other creative domains. The AI Drawing Partner is unique because it is both a co-creative agent that draws with the user (e.g. drawing lines, sketches, and images on a shared digital canvas), and a research platform that quantifies co-creation by coding the user and agent’s interactions, logging the action history of the user, and recording the drawing behavior of both the user and agent. The AI Drawing Partner is a freely accessible online platform \cite{AIdrawingPartner} that has two primary audiences: 1) co-creative AI researchers who can use it to conduct empirical studies, and 2) creative individuals (both novices and experts) seeking inspiration in their art-making process.

There are several approaches for evaluating co-creative AI systems \cite{kantosalo2020modalities,kantosalo2015interaction}. For example, the analysis can focus on the process, product, or individual \cite{karimi2018evaluating}. In this paper, techniques are presented for evaluating the co-creative process. The co-creative process includes interaction dynamics (e.g. turn-taking \cite{winston2017turn}, communication strategies \cite{rezwana2023designing}, modes of interaction \cite{davis2024creative}), cognitive dynamics (sense-making patterns \cite{de2007participatory}), collaboration dynamics (improvisational offers \cite{fuller2011shared}), and domain behaviors (e.g. drawing behaviors in this case). The AI Drawing partner utilizes the CCSM framework to quantify elements from each of these four categories to provide a systematic representation of the co-creative session. Additionally, the Four C model \cite{kaufman2009beyond} can be used to delineate the specific types of creative activities that co-creative systems \cite{rafner2023creativity} like the AI Drawing Partner aim to enhance, providing a structured way to evaluate and enhance these systems' impact on human creativity. The Four C model categorizes creativity into four distinct levels: Mini-c, Little-c, Pro-C, and Big-C, reflecting a spectrum from everyday creative actions to groundbreaking achievements \cite{kaufman2009beyond}.  The AI Drawing Partner primarily aims to enhances Mini-c and Little-c creativity. 

This paper begins by comparing and contrasting related co-creative drawing tools and the AI Drawing partner using co-creative framework for interaction design (COFI) \cite{rezwana2023designing}. Next, it describes the participatory sense-making and creative sense-making frameworks. Then, it introduces the co-creative sense-making framework. The AI Drawing Partner is described next, which utilizes the CCSM to automatically quantify interaction dynamics. There is then a case study presented of an expert artist interacting with the AI Drawing Partner for ten co-creative drawing sessions. The analysis is conducted on two groups of co-creative sessions: abstract and representational. Significant differences in interaction dynamics are found between the two groups.  The analysis includes showing the artworks produced, the quantified data from the AI Drawing Partner, the curves describing interaction dynamics (called creative sense-making curves), and a trend sequence classification process using stock market technical analysis. A discussion follows about how this approach can generalize to other co-creative AI systems. 

The primary contribution of this paper is presenting the AI Drawing Partner, which is a unique co-creative drawing agent and research platform that quantifies, models, and visualizes the co-creative process. The secondary contribution is presenting the domain independent CCSM framework and its quantification and analysis techniques. The tertiary contribution of this paper is demonstrating the new analysis technique for analyzing and modeling the interaction and collaboration dynamics of co-creative AI through a case study. The CCSM framework would enable a shared analysis technique that could advance the field of co-creative AI by enabling a common metric of comparison between co-creative experiences.

\section{Background}

\subsection{Maping the landscape of co-creative drawing tools with COFI}

Co-creative drawing systems feature a co-creative drawing agent that collaborates with the user on a creative product. The co-creative agent analyzes the user’s sketch contribution and generates a response to the user’s input. In some instances, sketch recognition is used to understand the user’s contribution and determine an appropriate response \cite{davis2016empirically}. A turn-taking process often emerges where the user and agent contribute content at different times \cite{winston2017turn}. Each turn can influence what comes next, and meaning and goals emerge dynamically as a result of the interaction between the user and agent \cite{davis2016co}. Co-creative drawing systems typically feature a shared canvas that both the user and co-creative agent contribute to, but some systems have separate canvases for the user and agent \cite{karimi2019deep}, where the agent’s contributions are meant to inspire the user, rather than directly contribute to the creative product. Co-creative drawing systems usually feature some form of generative AI model, such as recurrent neural networks (for generating the strokes of an object) \cite{wu2018sketchsegnet,sarvadevabhatla2016enabling} and convolutional neural networks (for recognizing sketched objects) \cite{davis2016co, li2020sketch}.

We can situate the AI Drawing Partner with respect to other co-creative systems in the field by considering the dimensions of the co-creative framework for interaction design (COFI) \cite{rezwana2023designing}. COFI was chosen to contextualize our work as it segments interaction between the collaborators (human and AI) and interaction with the shared product. This segmentation is critical, as the AI Drawing Partner features interactions between collaborators (e.g. voting, requesting sketches/images), as well as interaction with a shared product (e.g. a digital canvas that both the user and genAI can contribute to). Within the interaction between humans and genAI are collaboration style and communication style. Collaboration style has three dimensions: participation, task distribution, and timing. The AI Drawing partner has a participation style of both parallel and turn taking depending on how the user is interacting with the system. The user can draw at the same time as the agent or wait for the agent to finish its turn. The task distribution is defined as the same task (versus task divided) as the user and AI are collaborating on a shared canvas. The timing of the initiative is unplanned as the user specifies when to take a turn and when to wait. 

The communication style has three dimensions: human-to AI intentional communication, human-to-AI consequential communication, and AI-to-human communication. Of the 92 systems the authors analyzed with COFI, 82.6\% have no AI-to-human communication, and 68.5\% of human-to-AI communication is direct manipulation \cite{rezwana2023designing}. This finding indicates a significant gap in the research of co-creative AI systems, namely the communication channel between the user and agent. The human-to-AI communication channels in the AI Drawing Partner are voice (speech input for text-to-image generation), text (text input for text-to-image generation), and direct manipulation of the interface. There are no human-to-AI consequential communication channels in the AI Drawing Partner as it is not tracking gaze, biometric, embodied interactions, or facial expressions. The AI-to-human communication channels are text (speech bubble communicating the agent’s speech), visual (delivering sketches and images in communication), and embodied (the presence of a virtual character that is speaking). 

Here below we describe related human-AI co-creative drawing systems, comparing and contrasting them to the AI Drawing Partner using dimensions from the COFI framework. 

The Creative Sketching Partner \cite{karimi2019deep} is a co-creative design tool that supports conceptual shifts in the expert designer’s ideation process. The system uses a convolutional neural network (CNN) to recognize the user’s sketch and then matches that sketch with a similar sketch from a different object category. The system then contributes that sketch to the canvas to help the designer re-interpret their initial sketch and prompt a conceptual shift in the creative process. In contrast, the AI Drawing Partner either draws a new object, or changes the user's object into a new object. 

Oh et al., \cite{oh2018lead} present a co-creative drawing partner called DuetDraw that focuses on following the user and helping with the details of a task at hand. This tool uses a recurrent neural network (RNN) to predict how to complete an object. It can draw another object that the user previously drew. The system finds open space in the drawing to place its drawing contribution. The tool can also help the user with colorizing a sketch. \cite{ibarrola2023collaborative} present a context-aware co-creative design agent called CICADA. This system can help users complete objects they are drawing. As the user draws more on the canvas, the system’s model of the user’s sketch updates and the system dynamically responds based on the input from the user. The AI Drawing Partner has a 'Draw Object Together' mode where the system uses an RNN to complete the user's sketched object, similar to CICADA and DuetDraw. However, the AI Drawing Partner has several other modes of interaction, such as 'AI Mode' where the system chooses what drawing algorithm or AI model to use based on the user's preferences. 

Lin et al. \cite{lin2020your} demonstrate a co-creative sketching robot, Cobbie, that adds onto sketched ideas. The small robot moves around on the page to sketch with the user. The authors compare drawing with a co-creative robot to a virtual agent on a computer, and they found that engaging with the robot helped users ideate more effectively by provoking novel ideas. The AI Drawing Partner is similar in that it has a shared canvas where a co-creative agent is collaborating and contributing ideas.  

Recent work by \cite{abdellahi2020arny} presents an emotionally aware co-creative design agent called Arny. This agent recognizes facial expressions to determine when an individual is experiencing a given emotion. The emotion value is used as feedback to the system to determine what kind of contribution to make. The AI Drawing Partner uses positive and negative feedback in the form of voting to inform the system what kind of contribution the user likes and dislikes.

StoryDrawer \cite{zhang2021storydrawer} is a co-creative drawing agent that supports storytelling in children. The child uses voice input to tell parts of a story, and the system uses natural language processing to extract elements from the spoken input to sketch on a shared canvas. The system also has an IDEA button that enables the child to request the system to sketch an object. There is a sketch completion mode where the child scribbles and the system turns the scribble into an object. The authors found that StoryDrawer helped inspire children to tell more complex stories. The AI Drawing Partner has a similar feature as StoryDrawer where the user can draw the beginning of an object and the system will complete it. The AI Drawing Partner also has a voice input to the text-to-image generation AI model. 

Jansen and Sklar, 2021. \cite{jansen2021predicting} developed a computer vision approach for quantifying artistic interaction of traditional art media (e.g. pen and paper). Their AI model quantified when the artist was present and when they were drawing. They propose to use this system in conjunction with a co-creative robot that draws along with the user. These two features captured by Jansen et al’s system (e.g. presence and drawing) are also captured by the AI Drawing Partner system. However, we add an additional metric of ‘regulation’ to more fully describe the interaction dynamics of the creative process. In the case of pen and paper drawing, regulation would be moving the paper, moving the body to get different perspectives, and changing utensils. In the AI Drawing Partner, regulation is manipulating the interface (e.g. changing brush type, choosing fill tool) and communicating to the agent (e.g. requesting a sketch). 

SmartPaint \cite{sun2019smartpaint} is a co-creative painting application that uses a generative adversarial network to generate an image based on rough painted input. The system takes a single rough painting and creates detailed paintings based on the shapes and colors present in the source input. The primary difference between the AI Drawing Partner and SmartPaint is that the AI Drawing Partner is a turn-taking system that enables iterative collaboration between the user and co-creative agent. SmartPaint is a single-shot execution based on the user’s painted input. 

Drawcto \cite{deshpande2021drawcto} is a multi-agent co-creative system that collaborates with the user on non-representational artworks. The system chooses between using a rule-based algorithm to determine its contribution and a Sketch-RNN model. There is a dialog box where the agent provides an explanation of its actions. The system also has a stylization network that stylizes the images in the theme of a painting input. The AI Drawing Partner also chooses between a rule-based set of algorithms and the Sketch-RNN model. However, the system uses object recognition to determine when to use the Sketch-RNN model. The AI Drawing Partner also has a dialog system that explains its actions. Drawcto does not have text-to-image generation capabilities or voting feedback like the Drawing Apprentice does. Drawcto is also not a quantified co-creative AI system. 

Collabdraw \cite{fan2019collabdraw} uses a RNN to complete drawings of the user. The authors compared solo drawing to human-AI collaborative drawing and found that collaborative drawing contained as much semantic information (e.g. the number of strokes per object) as solo drawing. Collabdraw does not have a shared canvas that both the agent and user contribute to through time. It is a small input window meant for drawing an individual object at a time. The AI Drawing Partner, on the other hand, has a large shared canvas intended  to display an entire artwork co-created with the agent. 

Reframer \cite{lawton2023drawing} is a co-creative drawing agent that uses a prompt-based system to guide the co-creative agent's interaction. Reframer is based on the CICADA architecture for completing objects. The user can specify 'focus regions' of the drawing with different prompts to build up meaning in the artwork and constrain the system to provide region-based output. The authors explore how Reframer enables control through the text input, as well as emergence from the generative output of the system. The AI Drawing Partner does not feature global or regional prompt-based assistance, like Reframer, but it does employ image generation via text prompts. The AI Drawing Partner explores user control with communication (e.g. requesting sketches/images and positive/negative feedback) and drawing modes.

The AI Drawing Partner is inspired by the co-creative drawing systems reviewed here, and it most closely resembles the Drawing Apprentice \cite{davis2013human,davis2015enactive, davis2016co}. The Drawing Apprentice is a co-creative drawing agent that collaborates with the user in realtime on a shared canvas. The system has voting buttons and a creativity slider to provide feedback to the system about which creative mode to choose from when it generates its response. Later versions of the Drawing Apprentice feature object recognition. The system analyzes line input, categorizes it, and selects a semantically similar item to draw on the canvas. The system used a convolutional neural network (CNN) to achieve sketch-based object recognition. The AI Drawing Partner builds on the core interaction design of the Drawing Apprentice, but adds text-to-image generation and it is an example of a quantified co-creative AI system, whereas the Drawing Apprentice did not quantify the co-creation happening on the tool. 

The co-creative AI systems reviewed in this section all feature dynamic interaction between a user and co-creative agent in the domain of drawing (see Table \ref{tab:Comparisons} for a comparison of the functionality of co-creative drawing systems reviewed here.) However, none of the systems surveyed also collect, quantify, and visualize interaction dynamics. The AI Drawing Partner is unique in that it can contribute to the co-creative process in real time, as well as collect data, model it, and visualize the interaction dynamics of co-creation between the user and the AI Drawing Partner. 

\begin{table}
    \centering
    \begin{tabular}{|p{2cm}|p{1.5cm}|p{1.5cm}|p{1.5cm}|p{1.5cm}|p{1.5cm}|p{1.5cm}|p{1.5cm}|} \hline 
         &  Quantifies Co-Creation&  Text-to-Image Generation&  Sketch Recognition&  Shared Canvas&  Turn-taking&  Human-AI Communication& AI-Human Communication\\ \hline 
         AI Drawing Partner&  X&  X&  X&  X&  X&  X& X\\ \hline 
         Drawing Apprentice&  &  &  X&  X&  X&  X& X\\ \hline 
         Creative Sketching Partner&  &  &  X&  &  X&  X& \\ \hline 
         DuetDraw&  &  &  X&  X&  X&  X& X\\ \hline 
         CICADA&  &  &  X&  X&  X&  X& \\ \hline 
         Cobbie&  &  &  &  X&  X&  X& \\ \hline 
         Arny&  &  &  &  X&  X&  X& X\\ \hline 
         Story Drawer&  &  &  X&  X&  X&  X& \\ \hline 
         SmartPaint&  &  &  X&  X&  &  & \\ \hline 
 Collab Draw& & & X& X& X& &\\ \hline 
 Drawcto& & & X& X& X& &X\\ \hline 
 Reframer& & X& X& X& X& X&\\ \hline
    \end{tabular}
    \caption{Comparison of co-creative drawing system functionality.}
    \label{tab:Comparisons}
\end{table}

\subsection{Design Frameworks for Co-Creative AI
}Several design frameworks have been proposed for co-creative AI. The Co-Creative Framework for Interaction Design (COFI) splits interactions between interaction between collaborators and interaction with a shared product \cite{rezwana2023designing}. Within the interaction between collaborators are subcategories, such as communication style and collaboration style. Within the interaction with the shared product are subcategories, such as creative process and creative product. COFI is useful in considering the different dimensions of interaction design of co-creative AI systems. Guzdial \& Riedl \cite{guzdial2019interaction} proposed a framework for co-creative AI that focuses on different types of turn-taking in a co-creative system. They outline various configurations of co-creative systems where the AI has different roles in the collaboration and engages in different types of turn-taking. This framework is useful in designing the turn-taking interactions of co-creative AI systems. Muller and Weisz \cite{muller2022extending} proposed a framework called Collaborative Humans and AIs in which they focus on initiative and agency of the human and AI during co-creation. The framework is designed to support multiple human users and stakeholders during the collaboration. 

These frameworks can be utilized to inform the interaction design of co-creative AI systems. The CCSM, on the other hand, is designed to record quantitative data about the co-creative process. As a design guide, the CCSM suggests data collection categories and methods for acquiring that data. As an analytical framework, the CCSM defines types of data that can be collected during the co-creative process and a methodology for analyzing that data. The CCSM produces unique data that describes the temporal dimension of co-creation as a time-series dataset. This data can be used to understand the interaction dynamics and creative behavior of a co-creative session. While existing frameworks can be used to design and analyze co-creative AI systems, the CCSM can be used to quantify co-creation. This quantification provides additional insights into co-creation and can be used to aid in the explainability of the AI system. 

\section{Co-Creative AI Theory}
To understand the impact these AI experiences will have on human creativity, a theoretical framework is required to describe the human-AI interaction in this new paradigm of human-AI co-creativity. This paper argues the cognitive science theory of enaction provides  theoretical tools to investigate co-creation, such as sense-making \cite{thompson2009making}, participatory sense-making \cite{de2007participatory, torrance2011inter,fuchs2009enactive}, creative sense-making \cite{davis2017creative, deshpande2023observable}, and interaction dynamics \cite{fuchs2009enactive, auvray2012perceptual}. The cognitive science frameworks of participatory sense-making and creative sense-making can be combined with theories of improvisation \cite{fuller2011shared} to yield the co-creative sense-making framework (CCSM). 

The cognitive theory of enaction emphasizes how meaning is built through exploratory interaction with the environment and others within that environment \cite{varela2017embodied}. Enactivists propose sense-making to be the cornerstone of cognition \cite{thompson2009making}, which is a process of interacting with the environment to cast a web of meaning relative to the sensorimotor feedback loops engaged during the interaction \cite{de2007participatory}. Participatory sense-making is a cognitive framework that delves deeper into the social dimensions of sense-making, where two or more agents coordinate both their actions and the broader interactional context \cite{de2007participatory, torrance2011inter,cuffari2015participatory}. Creative sense-making \cite{davis2017creative, davis2017quantifying, davis2024creative} is another cognitive framework that examines sense-making in the context of a creative activity, i.e. the conversation with the materials \cite{schon1992designing} that emerges in crafting, designing, and artmaking. Creative sense-making proposes modes of interaction a user engages in as they co-create with an agent. These interaction modes are continuously coded to provide quantified interaction dynamic data modeling the co-creative experience. The CCSM combines the previously described frameworks, providing a cognitive framework and data collection schema for co-creative systems to implement. These quantified co-creative systems automatically code the interactions of collaborators and quantify interaction, collaboration, cognitive, and domain  dynamics.

\subsection{Participatory Sense-Making}
One core tenet of the cognitive science theory of enaction is that meaning is built through interaction with the environment \cite{di2010horizons}. Agents engage in a process called sense-making, or regulating their interaction based on coordination and feedback from the environment to maintain their autonomous identity \cite{de2007participatory}. Meaning emerges during sense-making while agents cast a web of significance \cite{di2010horizons} onto the environment through their interactions and the results of those actions in sensorimotor loops \cite{o2001sensorimotor, noe2004action, di2017sensorimotor, degenaar2017sensorimotor}. These sensorimotor loops develop into sensorimotor contingencies, such that sensory data is understood in terms of bodily constraints and capabilities. In this way, cognition is embodied \cite{clark1999embodied, varela2017embodied}, embedded, and situated in the environment \cite{Robbins08, suchman2007human}.

When multiple agents are involved in sense-making, it becomes more complex as a new domain of relational dynamics emerges in a process called participatory sense-making: “...the coordination of intentional activity in interaction, whereby individual sense-making processes are affected, and new domains of social sense-making can be generated that were not available to each individual on their own” \cite{de2007participatory}. This coordination can be directed toward contributing to the interaction (interactional coordination) or maintaining the interaction (functional coordination) \cite{de2007participatory}. This distinction has been applied to co-creation to denote two distinct processes in co-creation: interaction between collaborators, and interactions with the shared product \cite{kellas2014rating, rezwana2023designing}. 

Participatory sense-making (PSM) emerges when collaborators are co-regulating both interactional sense-making processes \cite{rezwana2023designing}. For example, co-regulation occurs when collaborators are communicating both through creative actions and feedback to their partner. During participatory sense-making, interactions are coupled such that each subsequent interaction is influenced by what came before \cite{de2007participatory}. “This coupling has a depth (the number of turns), a duration (total time spent coupled), an initiator (the individual that began the coupling), and a decoupler (the individual that ended the coupling)" \cite{davis2024creative}. The AI Drawing Partner quantifies these elements of interaction coupling.

\subsection{Creative Sense-Making}
The creative sense-making (CSM) framework describes creative cognition as consisting of sense-making cycles where the cognitive agent interacts with its environment to iteratively build meaning and explore its creative product \cite{davis2017creative}. There are two states of cognition in this framework, engaging in sense-making to understand how something in the environment works (e.g. inspecting or experimentally interacting), and interacting effectively with that system in the environment once the agent has made sense of the system in the environment (e.g. creative flow \cite{csikszentmihalyi1990flow}). When the agent is actively engaged in sense-making, this mode of cognition is referred to as unclamped cognition.  When the agent knows what to do and is effectively interacting with the environment, this mode of cognition is referred to as clamped cognition. Creative sense-making is characterized by fluctuations between clamped and unclamped cognition. Cognition can unclamp in two ways that correspond to the types of sense-making identified by \cite{de2007participatory}: functional unclamp and interactional unclamp. 

A functional unclamp event is where the individual is modifying or regulating the manner in which they are interacting with the environment and other agents within the environment, such as exploratory and metacognitive actions. 
Defined as the reflection and management of one's own cognitive actions—essentially, cognition about cognition \cite{norman2019metacognition}, —metacognition has been studied extensively in psychology, and widely accepted as playing an important role in creativity \cite{armbruster1989metacognition,pesut1990creative}. Metacognitive activities involve the adaptive monitoring and regulation of cognitive functions, such as evaluation of ideas and task performance, or developing and selecting strategies for tasks \cite{lebuda2023systematic}.  The creative metacognition framework divides up phases of an interaction to include pre-task or task preparation (e.g. process task instructions), performance stage during task (e.g. core task such as idea generation), and post task (e.g. presentation/implementation of solution) \cite{lebuda2023systematic}.

The different phases of the creative metacognition framework can be partially mapped to clamped/unclamped cognition. During task preparation, the individual would be unclamped and making sense of the task. During the performance stage, there would be cycles of clamped cognition, where the individual is executing  actions, and unclamped cognition, where the individual is exploring, evaluating, and self-regulating. Creative sense-making enables the researcher to make fine-grained distinctions about the interaction dynamics of sense-making in a creative task, whereas the creative metacognition framework identifies broader phases of creativity.

In the domain of co-creative drawing, a functional unclamp would be exploring the interface of the tool, editing the paint brush, rotating the page, zooming, as well as any form of communication to the AI. Interactional unclamp is when the individual disengages from the interaction, such as pausing, hesitation, and non-action. Taking a step back from an artwork to get a global view would be an example of an interactional unclamp. The interaction modes and their corresponding cognitive modes are depicted in Table \ref{tab:interactionMode}.

\begin{table}
    \centering
    \begin{tabular}{|c|c|c|c|} \hline 
         Interaction Mode&  Example&  Cognitive Mode& Coding Value\\ \hline 
         Communicate to the AI&  Providing positive/negative feedback to the agent&  Functional Unclamp& 1\\ \hline 
         Manipulate Interface&  Edit brush, activate settings panel, undo/redo&  Partial Functional Unclamp& .5\\ \hline 
         Waiting&  Pausing, hesitation, and non-action&  Interactional Unclamp& 0\\ \hline 
         Execute Artistic Action&  Pen on paper drawing&  Clamped& -1\\ \hline
    \end{tabular}
    \caption{The cognitive modes of the CSM mapped onto the interaction modes and coded value.}
    \label{tab:interactionMode}
\end{table}

Creative cognition can then be characterized by sense-making cycles featuring a cycle between clamped and unclamped cognition \cite{davis2015enactive}. Conversely, everyday cognition can be characterized as clamped cognition with minor unclamp events when something surprising happens \cite{davis2015enactive}. The CSM framework is used to quantify these cognitive fluctuations and define sense-making patterns and trends that characterize creativity. In particular, it has been used to code the interaction dynamics of pretend play \cite{davis2017creative}, improvisational dance \cite{deshpande2023observable}, and co-creative drawing \cite{davis2017creative,davis2017quantifying}. When applied to co-creative AI, there are four interaction modes formed by the modes of cognition: fluidly executing actions (clamped cognition), manipulating the interface of the system (partial functional unclamp), communicating to the AI (functional unclamp), and waiting, pausing, hesitation, or non-action (interactional unclamp) \cite{davis2024creative}. These interaction modes are given a value (communicate = 1; manipulate interface = .5; wait = 0; execute = -1)  and continuously coded through time. The coded values represent a continuum from clamped cognition (-1) to unclamped cognition (1). The polar values of the continuum are arbitrary as unclamped cognition could be defined as 1. Waiting is coded as 0 so waiting and non-action do not influence the direction of the trend. Waiting is a balanced state between clamped and unclamped cognition. These coded values produce two kinds of data: the raw interaction mode count, and the cumulative sum of the raw interaction mode count, which forms the creative sense-making curve, as shown in Figure \ref{CombinedFig}. The creative sense-making curve visually depicts the interaction dynamics trends and patterns in the creative session. 

\subsection{Co-Creative Sense-Making Framework}
The co-creative sense-making framework was developed  in order to model the interaction and collaboration dynamics of co-creative AI systems. It is a novel approach that quantifies elements of co-creation in situ. The data it generates enables comparisons between the interaction of the user and co-creative agent, as well as between participants in an empirical study. It is a domain independent framework that can enable cross-domain comparisons of co-creative behavior. 

The co-creative sense-making framework is based on the cognitive science frameworks of participatory sense-making \cite{de2007participatory}, creative sense-making \cite{davis2017creative}, and theories of improvisational creativity \cite{fuller2011shared}. The CCSM combines the theoretical framework of PSM, such as the concept of interaction dynamics, with the analytical framework of CSM to arrive at a framework that quantifies the sense-making processes of co-creation. 

\begin{figure}[h]
  \centering
  \includegraphics[width=\linewidth]{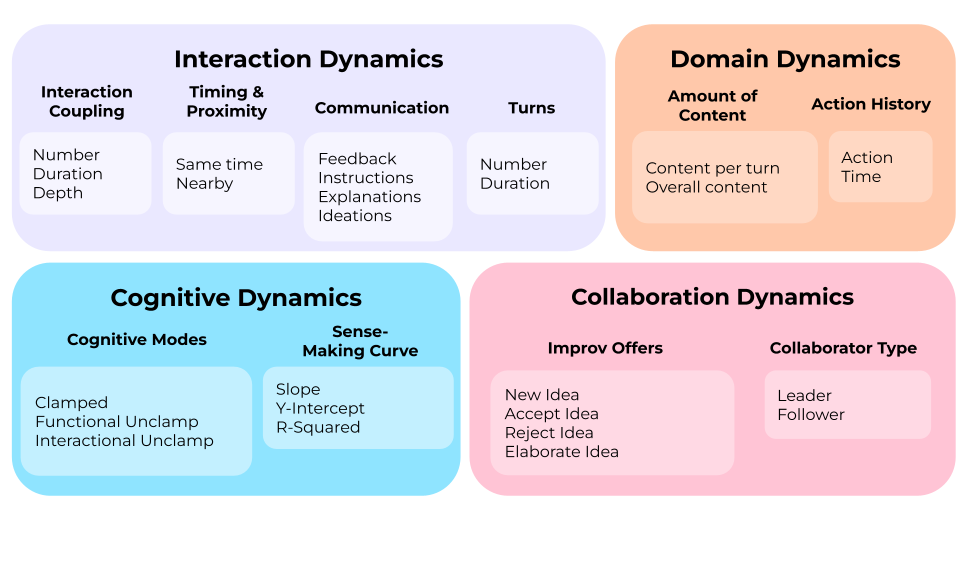}
  \caption{The Co-Creative Sense-Making Framework. Each category has a number of features that can be quantified. This categorization serves as a data collection schema for quantified co-creative AI systems.}
  \Description{}
  \label{Framework}
\end{figure}

Figure \ref{Framework} shows the CCSM framework, which is segmented into four main categories. Each category has subcategories that each have features that can be quantified in a co-creative system. This framework can be used to design the data collection scheme for co-creative systems. Collecting this type of data can help researchers understand how co-creative AI systems impact the creativity of users. In particular, it quantifies and models the interaction and collaboration dynamics of co-creativity. 

In this paper, we use the CCSM to evaluate the AI Drawing Partner system. Data was collected from 10 co-creative drawing sessions. The CCSM enabled a comparative analysis of abstract and representational art-making with the AI Drawing Partner. Based on the findings, the framework can be generalized to evaluate other co-creative systems. 

The CCSM has four categories of data that are recorded and stored from a co-creative interaction: cognitive dynamics, interaction dynamics, collaboration dynamics, and domain behaviors. Each of these categories are elaborated on below. 

\subsubsection{Cognitive Dynamics}
\begin{figure}[h]
  \centering
  \includegraphics[width=\linewidth]{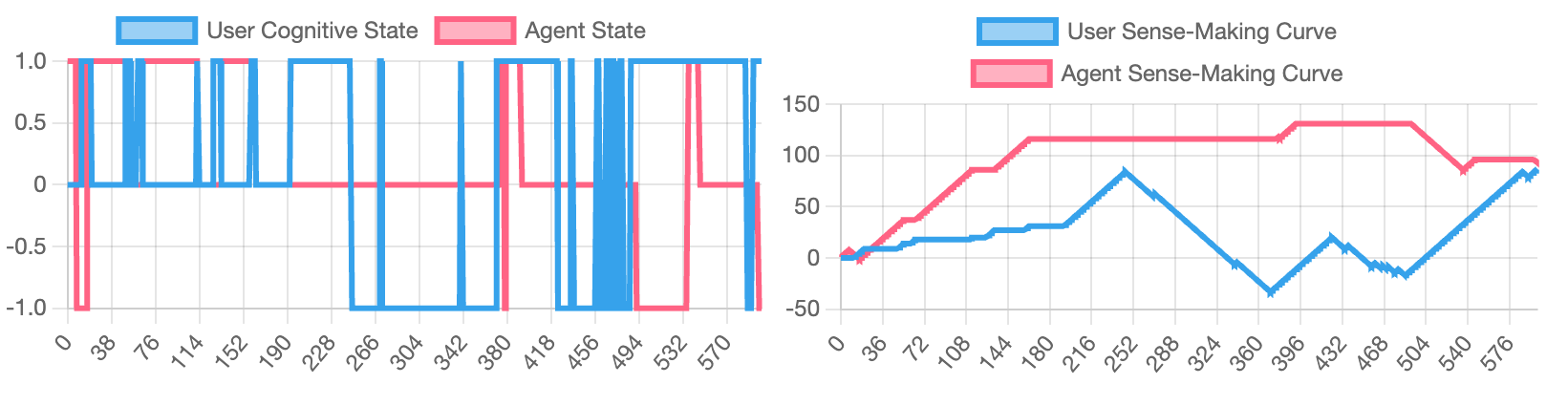}
  \caption{Raw coded values (left) and creative sense-making curve (right) for a five-minute co-creative drawing session.}
  \Description{}
  \label{CombinedFig}
\end{figure}

Cognitive dynamics relate to the mode of cognition (clamped/unclamped) an individual is engaged in through time. The CSM coding convention is used to produce the raw interaction dynamic data (coded interaction modes) and the creative sense-making curve, as shown in Figure \ref{CombinedFig}.

 The CSM curve depicts the interaction dynamic trends. When the curve rises, the individual is regulating their interaction (communication and interface manipulation). When the curve falls, the individual is fluidly executing actions (e.g. drawing). When the curve is horizontal, that means the individual is waiting. Waiting can be both waiting for the agent to take its turn, and pausing and hesitation during a turn (e.g. pausing to think). Waiting is defined as any time the pen is not down, and the user is not manipulating the interface or communicating to the agent. The slope of the CSM curve provides a quantitative value describing the degree to which the individual was regulating or executing during the creative session. A linear regression can be applied to the CSM curve to attain an r-squared value, which describes how linear the CSM curve is. More linearity would correspond to more consistent interactions during the creative session. The analysis then classifies trends in the data using stock market technical analysis in order to reveal the sequential order of events in the dataset. Here, buy, sell, and hold classifications are determined in the data. For the purpose of the analysis, buy would correspond to regulate, sell would correspond to execute, and hold would correspond to wait. These trends are visualized to identify sequential events and enable a comparative analysis between the sessions.

\subsubsection{Interaction Dynamics}
Interaction dynamics is a term used by the PSM framework to describe the manner in which social coordination occurs, including turn-taking \cite{cuffari2015participatory,ikegami2007turn}, communication strategies \cite{de2007participatory}, and interaction coupling \cite{auvray2012perceptual}. Turn-taking can be analyzed by counting the number of turns and examining the length of turns \cite{davis2024creative}. Communication is a critical component of co-creation as it is used to negotiate meaning \cite{rezwana2023designing}. There are several types of communication present in co-creation, such as feedback \cite{davis2015drawing}, explanations \cite{ehsan2022human}, instructions \cite{rezwana2021creative}, and ideations \cite{karimi2020creative}. An interaction coupling occurs when successive turns are mutually influential such that there is some structural correspondence between the content of the turns \cite{de2007participatory}. When an interaction coupling occurs, it has an initiator (i.e the collaborator that began the coupling), decoupler (i.e. the collaborator that ended the coupling), number of turns, and amount of content per turn.

\subsubsection{Collaboration Dynamics}
In the CCSM, collaboration dynamics uses improvisation theory to analyze who introduced new content, whether that content was accepted or rejected by the partner, and elaborations upon established ideas. In improv, introducing new content into the improvisation is referred to as making an offer \cite{fuller2011shared}. This offer can be accepted or rejected. If it is accepted, the improv actor engages in a ‘yes, and’ behavior that acknowledges the validity of their partner’s contribution and adds upon it \cite{fuller2011shared}. Once an offer is accepted, it can be elaborated by the participants. In the CCSM, offers are referred to as introducing new ideas, and they can be accepted, rejected, and elaborated depending on the reaction of the partner. 

\subsubsection{Domain Dynamics}
\begin{figure}
  \centering
\includegraphics[width=0.7\textwidth]{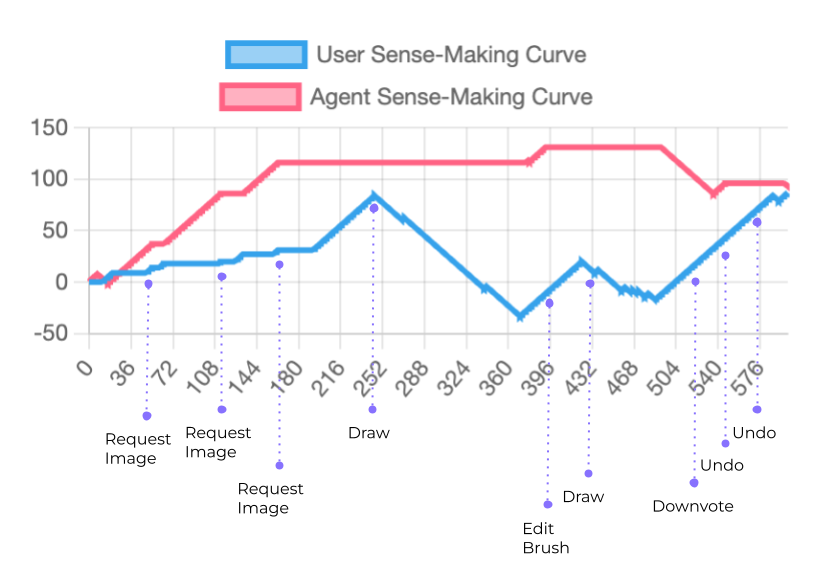}
  \caption{Action history overlaid onto the CSM curve for the user in Session 6.}
  \Description{}
  \label{ActionHistory}
\end{figure}

Domain behavior relates to what type of creative actions occurred and the nature of the content that was produced, relating to the creative product \cite{karimi2018evaluating}. It is possible for co-creative systems to record the interactions of the user to understand more about domain behavior. In the domain of co-creative drawing, domain behaviors would include: draw, fill, erase, undo, redo, smudge, clear canvas, choose color, choose line width, provide feedback, request the system to sketch, request the system to generate image, and teach the system a new object. The domain behaviors can be overlaid onto the CSM curve to analyze what type of interaction was present for a given interaction dynamic trend, as shown in Figure \ref{ActionHistory}.


Co-creative systems can also quantify the amount of creative content produced, such as the total amount of content produced (number of lines drawn), the features of that content, such as the length of lines drawn on the canvas, how much content was produced per turn (lines per turn), as well as the speed with which contributions were made (drawing speed).

\section{AI Drawing Partner System Design}
\begin{figure}[h]
  \centering
  \includegraphics[width=\linewidth]{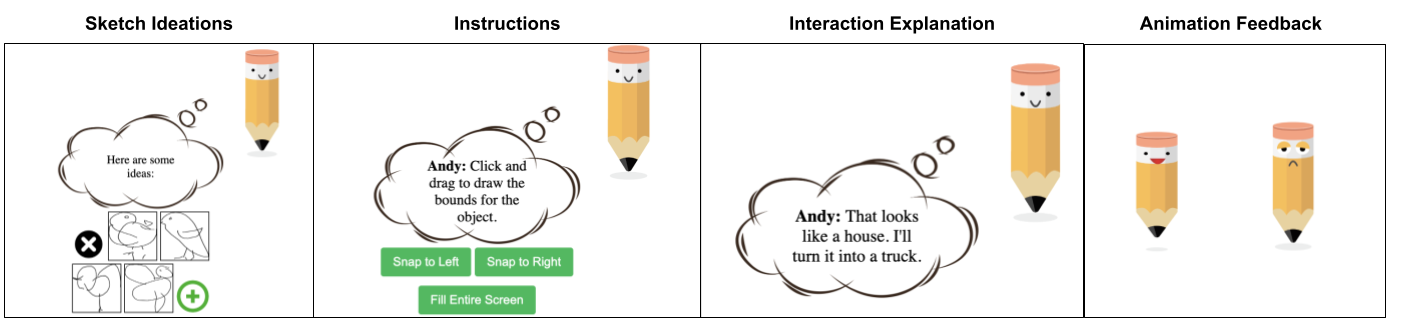}
  \caption{The AI-Human communication channels in the AI Drawing Partner.}
  \Description{}
  \label{CommChannels}
\end{figure}

The AI Drawing Partner is a web-based co-creative drawing system that collaborates with the user in real time on a shared canvas (see Figure \ref{CommChannels}). The tool is publicly available at: \cite{AIdrawingPartner}.  The defining feature that sets the AI Drawing Partner apart from other co-creative drawing systems and co-creative AI systems is that the AI Drawing Partner both collaborates with the user and quantifies the interaction dynamics of co-creation. There are four main components to the interface: the menu, the paint menu, the chat interface, and voting buttons. The menu is located at the top of the interface and contains functionality for controlling the collaboration, such as the settings panel, where users can choose the collaboration mode, which is displayed at the top left corner. The paint menu is located on the left side and includes functionality for editing the brush and manipulating images and layers. The chat interface is located at the bottom left, and it includes a dropdown selector for requesting the system to sketch objects and a text input box and voice input mechanism for prompting the system to generate an image. The voting buttons are located on the right side, and they control feedback to the system. This feedback is used to train the system’s algorithms to align to the user’s preferences through time.

\begin{figure}
  \centering
  \includegraphics[width=\linewidth]{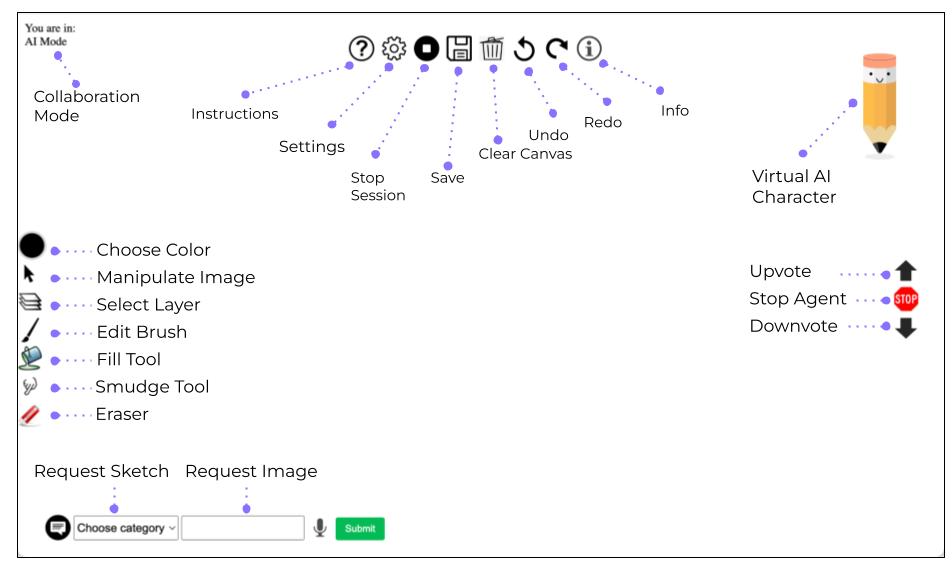}
  \caption{AI Drawing Partner interface with annotated functionality.}
  \Description{}
  \label{DrawingApprentice}
\end{figure}

The collaborative drawing interaction is loosely turn based, but the user can draw while the agent is drawing, which makes simultaneous collaboration possible. The interaction design of the AI Drawing Partner can be broken down into the two interactional sense-making categories identified by \cite{kellas2014rating}: interactions between partners and interaction with the shared product, as shown in Figure \ref{DrawingApprentice}. The interaction between partners consists of human-to-AI communication and AI-to-human communication. The human-to-AI communication channel in the AI Drawing Partner has three dimensions: providing positive and negative feedback with voting, instructing the system to perform an action (e.g. selecting drawing modes, requesting the system to draw a bird), and providing a text prompt for the system to generate an image. The AI-to-human communication channel has four dimensions: a speech bubble from the agent describing what action it is going to take, positive or negative animations in response to feedback from the user, thumbnail sketch ideations, and instructions for the user to follow in the interaction (see Figure \ref{CommChannels}). The AI agent engages in a jumping animation after a positive feedback event. When users provides negative feedback, the agent immediately stops what it is drawing, and displays a frown face to denote the reception of the negative reinforcement.

The user and AI processes are a subset of the mixed-initiative creative interfaces categories for the flow of the creative process: ideate, constrain, produce, select, assess, and adapt \cite{deterding2017mixed}. In the current version of the AI Drawing Partner, the user and AI are ideating, producing, suggesting, selecting, and adapting. New processes, such as task-based reasoning and aesthetic evaluation, can be added to the system to enable the user and AI to constrain the artistic task and assess the contributions that have been added to the shared creative product.

\begin{figure}[h]
  \centering
  \includegraphics[width=\linewidth]{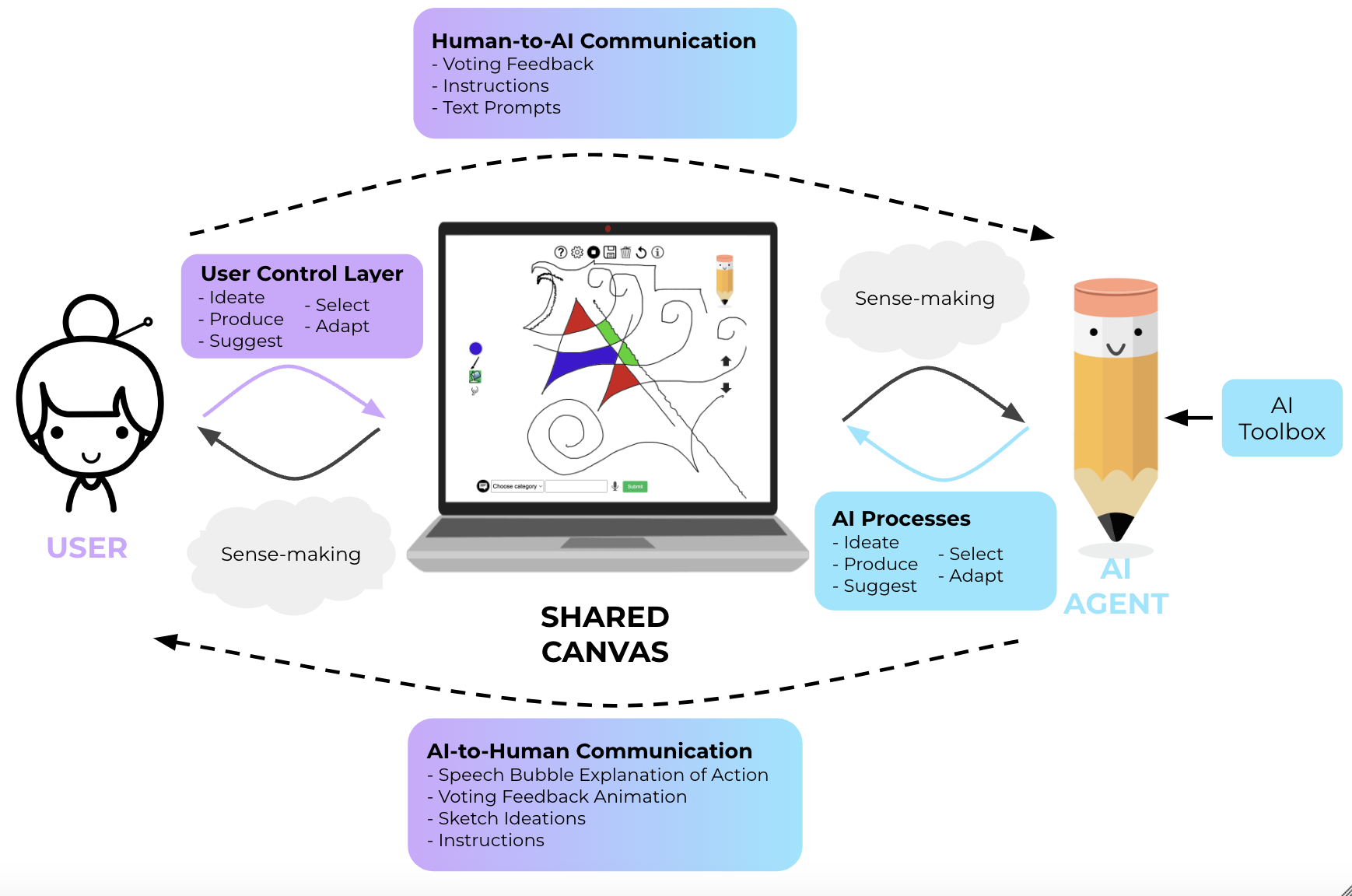}
  \caption{Interaction design diagram of AI Drawing Partner system. The user is situated on the left, while the AI agent is situated on the right. The interactions are divided into two categories: interactions with the shared canvas and interactions between collaborators (e.g. communication). The AI Toolbox (shown in Table 2) informs the creative decisions of the AI agent. }
  \Description{}
  \label{DesignDiagram}
\end{figure}

The user can interact with the shared product in several ways, including: ideating new content, producing new contributions, suggesting contributions for the system to make, selecting which sketch the agent should contribute, and adapting the agent’s contributions to their own. The agent can interact with the shared product in a similar capacity, by ideating sketches, producing new contributions (e.g. new sketches or images), suggesting new sketch objects to add, selecting sketch objects from the database to add, and adapting lines to the user’s lines (e.g. scaling, translating, rotating, adding noise). The system utilizes the AI Toolbox to construct its creative contributions. New AI models can be added to the toolbox to enable new types of interactions in the system. 

\begin{table}
  \centering
  \caption{AI Toolbox}
  \label{tab:AIToolbox}
  \begin{tabular}{p{0.3\linewidth}p{0.3\linewidth}p{0.3\linewidth}}
    \toprule
    Algorithm & Interaction & User Input \\
    \midrule
    Sketch-RNN (Ha \& Eck, 2017) & Transform Object, Generate Object & Lines, Object Selection \\
    CNN Image Stylization (Huang \& Belongie, 2017) & Transform Entire Drawing & Input Image, Style Strength \\
    Stable Diffusion (Rombach et al., 2022) & Text-to-Image Generation & Text Prompt \\
    Reactive Algorithms & Mimic & Lines, Feedback \\
    CNN Sketch Recognition & Sense-Making & Lines \\
    \bottomrule
  \end{tabular}
\end{table}

\begin{figure}[h]
  \centering
  \includegraphics[width=\linewidth]{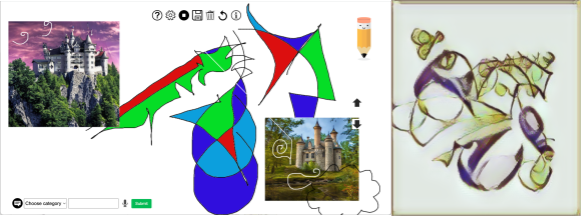}
  \caption{Demonstrating interactive text-to-image generation on a shared canvas (left) and image stylization (right).}
  \Description{}
  \label{ImageStyl}
\end{figure}

\section{System Architecture}

\begin{figure}[h]
  \centering
  \includegraphics[width=\linewidth]{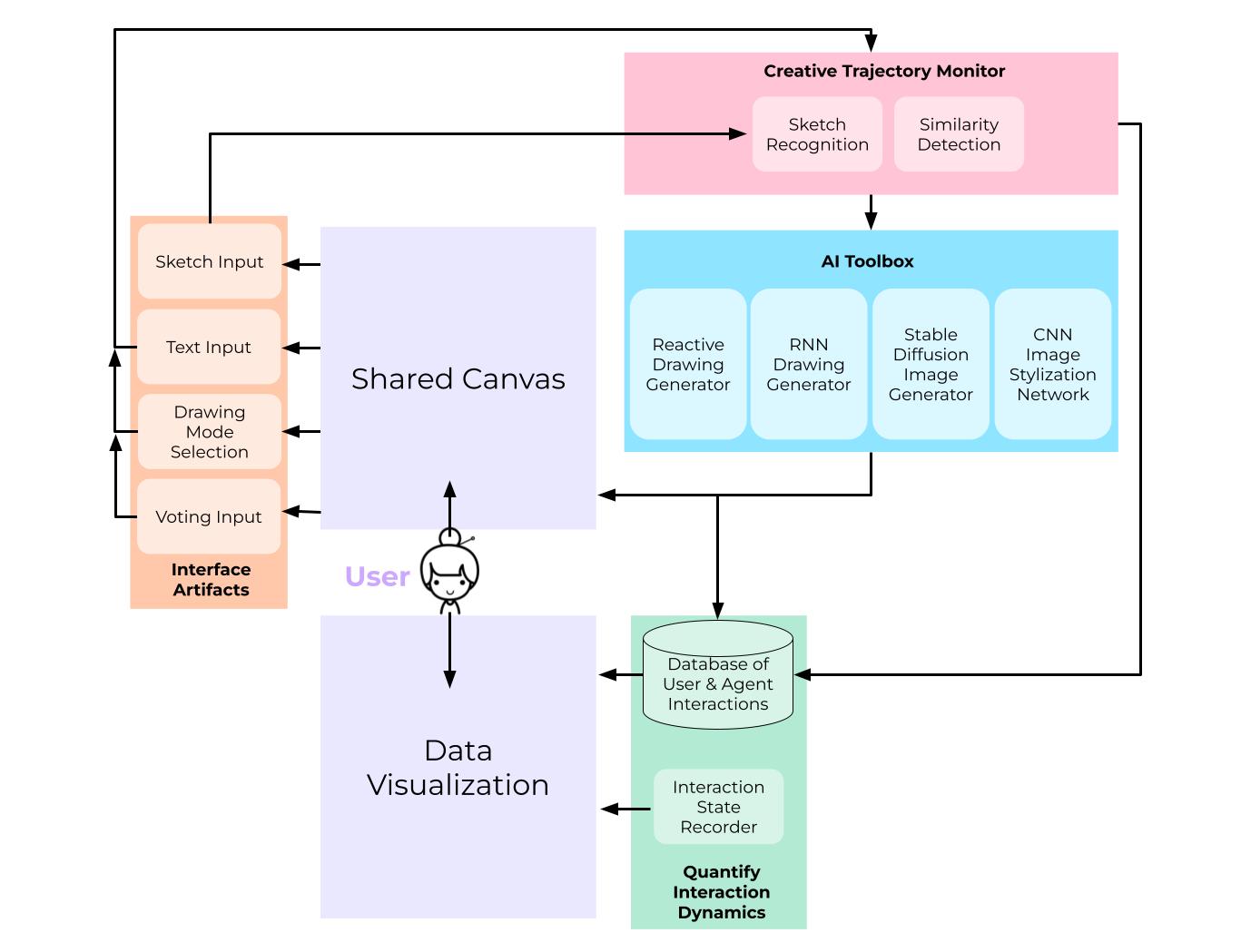}
  \caption{The AI Drawing Partner system architecture. The interface artifacts are all the user input into the system. The creative trajectory monitor takes user feedback and sketch input to direct the AI algorithms in the AI toolbox. The system’s generated response is sent to the canvas and database to quantify interaction dynamics. The data visualization shows the quantified results of the co-creative session.}
  \Description{}
  \label{SysArch}
\end{figure}

The shared canvas element is where user interaction occurs. It is the source of input for the system. There are four distinct sources of human input (communication with the AI): voting data, sketch input, text input, and drawing mode selection. The voting data is binary positive or negative feedback. It is used to improve the selection of the reactive algorithms of the system. When users provide positive feedback, the probability that that reactive algorithm increases linearly. 

The sketch input is a series of x,y points that form a line. There can be multiple lines per turn. Each turn is determined by a three second timer after the user stops drawing. Sketch input is saved and reused by the drawing algorithms for generating lines later in the drawing. User input populates a local database from which the system randomly selects in some of its reactive algorithms. For example, when the system is extending the user’s line, the line segment it chooses to extend the line with might be one of the user’s previously drawn lines. 

The sketched input lines are sent to a similarity detection algorithm which uses Frechet distance \cite{alt1995computing}, which compares the similarity between two polygonal curves, to compare all the lines from the current turn to the lines from the previous turn. If there is a significant enough similarity, a flag is thrown in the system to denote the fact that the user is engaged in a repetitive pattern-like behavior. In this case, the extend algorithm is selected and the same input is attached to the end of the user’s last input line. 

The text input comes from the text chat box in the lower left corner of the user interface. Here, the user can prompt the system to generate an image based on the text they input into the system. The system calls the stability.ai API to request an image from the platform, which uses the Stable Diffusion AI model \cite{rombach2022high}. When the user makes a request of the system, such as drawing an object or generating an image, the output presents an option to the user about where to place that data and what size it should be. This layer of interactivity helps the user have more creative control. 

\begin{table}
  \caption{The collaboration modes in the AI Drawing Partner}
  \label{tab:CollabModes}
  \begin{tabular}{p{0.25\linewidth}p{0.7\linewidth}}
    \toprule
    Collaboration Mode & Description \\
    \midrule
    AI Mode & System chooses which drawing algorithm to use based on user preferences (voting). \\
    Draw Together & The user chooses an object to draw with the system, then draws the first line. The system finishes the object. \\
    Extend & The system extends the user’s last contribution. \\
    Mimic & The system imitates the user’s last contribution by rotating, scaling, and translating it. \\
    Shade & The system adds shading lines to the user’s last contribution. \\
    Predictive Drawing & The system displays the lines it will draw in the next turn before it draws them. The user votes up for the system to draw the lines, and votes down for the system to generate a new idea. \\
    Style & The user selects a style that the system will use to respond with. There are two styles by default: artistic and neural. In artistic style, the agent responds with a set of stored artistic lines. In neural style, the agent responds with an AI model trained on the artistic lines. \\
    Teach Agent Object & The user draws an object on the canvas, then tells the system what object it was. The object is stored in the agent’s knowledge, and it can draw it when requested. \\
    Teach Agent Style & The user draws a series of lines on the canvas, then tells the system the name for the style. The style can then be selected, and the agent will utilize the lines the user input into the style to generate its responses. \\
    \bottomrule
\end{tabular}
\end{table}

The drawing mode, as shown in Table \ref{tab:CollabModes}, determines which type of drawing algorithms the system will use. In AI Mode, the system selects which algorithm to use based on what it has learned from the user. In Draw Object Together mode, the user and system can draw together. The user inputs a starting line or series of lines, and the RNN network completes the object. In the extend mode, the system extends lines with either the same line the user input, or another line from a previous point in the drawing. Mimic mode translates and rotates the user’s input to repeat patterns they have drawn. The select style mode has multiple saved styles to select from. The first is the artist style, which uses ~150 lines that an artist (the first author) drew with the system. The system randomly selects among those lines to respond to the user’s contribution when this mode is selected. The other style is neural. The neural style uses a convolutional neural network trained on the database of artist lines to generate a response. 

To recognize what the user drew during their turn, a bounding box is created around the lines they drew. A new canvas is created with the size of that bounding box, and the lines are drawn within the new canvas. The canvas is then sent to the CNN sketch recognition engine, which uses the ml5 object recognition model \cite{ml5}. It returns several results with their corresponding confidence values. If the confidence of a potentially recognized label for the sketch input is above the confidence threshold (which is set to 30\% by default), then the system will respond as if the user drew an object. In this case, the system checks the similarity of the words between the recognized object and all the objects in the database computed using TensorflowJS word embeddings. If the word is present in the database, the system will draw a version of that object. Otherwise, it will select the word with the least semantic distance in the database and draw it in an open space on the canvas. 

The creative trajectory monitor (CTM) takes in data from the similarity detector and sketch recognition and makes a determination about what mode the drawing agent should be in. This mode determination is influenced by the voting behaviors of the user. The CTM chooses which exact reactive algorithm (if any) the system will use during its next turn. 

The reactive drawing generator takes input lines and transforms them to produce an output line. This retains some of the human creative qualities of the lines, while introducing novelty in the placement and scale of the lines. The generator can add noise to existing lines, translate, rotate, scale, and combine those operations to achieve a variety of effects.  

The Sketch-RNN \cite{ha2017neural} drawing generator consists of a database of about ~110 objects  (e.g. ant, angel, backpack, butterfly, basket, bear, bee) the system knows how to draw. The system works by selecting a category, then the algorithms generate examples of that category based on a model it has learned from seeing many thousands or millions of examples. It generates stroke level data that can be cleaned and used in the AI Drawing Partner to draw sketches of common objects. 

The Stable Diffusion image generator uses the Stability.ai API for the Stable Diffusion XL Base 1.0 model \cite{rombach2022high}. The API accepts a string of text from the user via a text input box and returns a single image representing the semantic content of the input string. This image is displayed as a thumbnail to the user to select. After selecting the thumbnail, the user is prompted to provide instructions about where to place the image and the size of the image. 

\section{Case Study}
To demonstrate the CCSM analysis process, a case study was conducted by an expert artist (the first author) with the AI Drawing Partner. This is not a user study evaluation, but rather a case study to demonstrate applying the CCSM analysis technique to a set of data. To utilize the CCSM, two groups of data are required to determine whether there are any statistical differences in creative behavior between the groups. In a user study, these groups could represent different skill levels (e.g. novice vs. experts). However, for the purposes of the case study, the sessions were segmented into abstract and representational image generation to determine whether the analytical method can determine any differences in interaction and collaboration between the distinctly different creative strategies. Each session lasted five minutes. Five minutes was selected as the duration as that is approximately how long it takes to fill up the canvas with lines in an abstract drawing based on experimentation. In the abstract group, the user was mostly sketching, as the user has skill in abstract drawing. As a result of this sketching behavior, the system utilized the reactive algorithms. In the representational sessions, the user utilized the generative AI algorithms in the system, such as the Sketch-RNN model for generating sketches, and the Stable Diffusion model for generating images, as the user is not skilled in representational drawing and they were using the system to assist them. 

The research questions and hypotheses are as follows: 

\begin{itemize}
    \item \textbf{RQ1}: How does communication change in abstract and representational art making?
    \begin{itemize}
        \item \textbf{Hypothesis 1}: There will be more communication in representational art given the user will be communicating their artistic intentions to the system more, e.g. asking the system to draw an object.
    \end{itemize}
    \item \textbf{RQ2}: How does creative behavior differ in abstract versus representational art making?
    \begin{itemize}
        \item \textbf{Hypothesis 2}: There will be more drawing actions and a greater quantity of lines produced in the abstract group due to lack of communication (e.g. requesting sketches/images) and focus on drawing, leading to a decreased slope in the CSM curve and a lower average coded value.
    \end{itemize}
    \item \textbf{RQ3}: What are the differences in interaction dynamics between abstract and representational art making?
    \begin{itemize}
        \item Hypothesis 3: There will be more fluid turn-taking and more turns in general in the abstract condition due to the nature of the real-time collaboration occurring in the abstract condition.
    \end{itemize}
\end{itemize}

\begin{table}
 \caption{Artworks generated from the abstract co-creative sessions (1-5 in the left column) and representational co-creative sessions (6-10 in the right column)}
  \label{tab:Artworks}
  \begin{tabular}{cc}
    \toprule
    Abstract & Representational \\
    \midrule
    1 \includegraphics[width=0.4\linewidth]{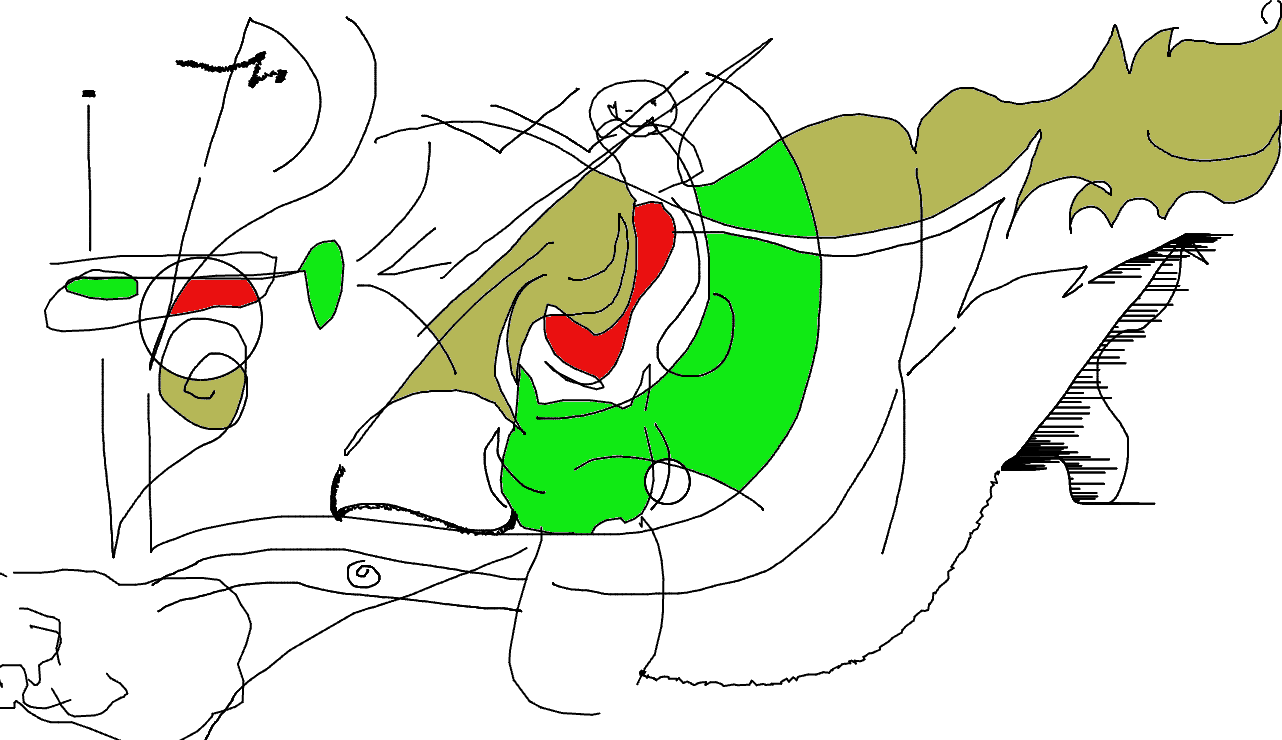} & 6 \includegraphics[width=0.4\linewidth] {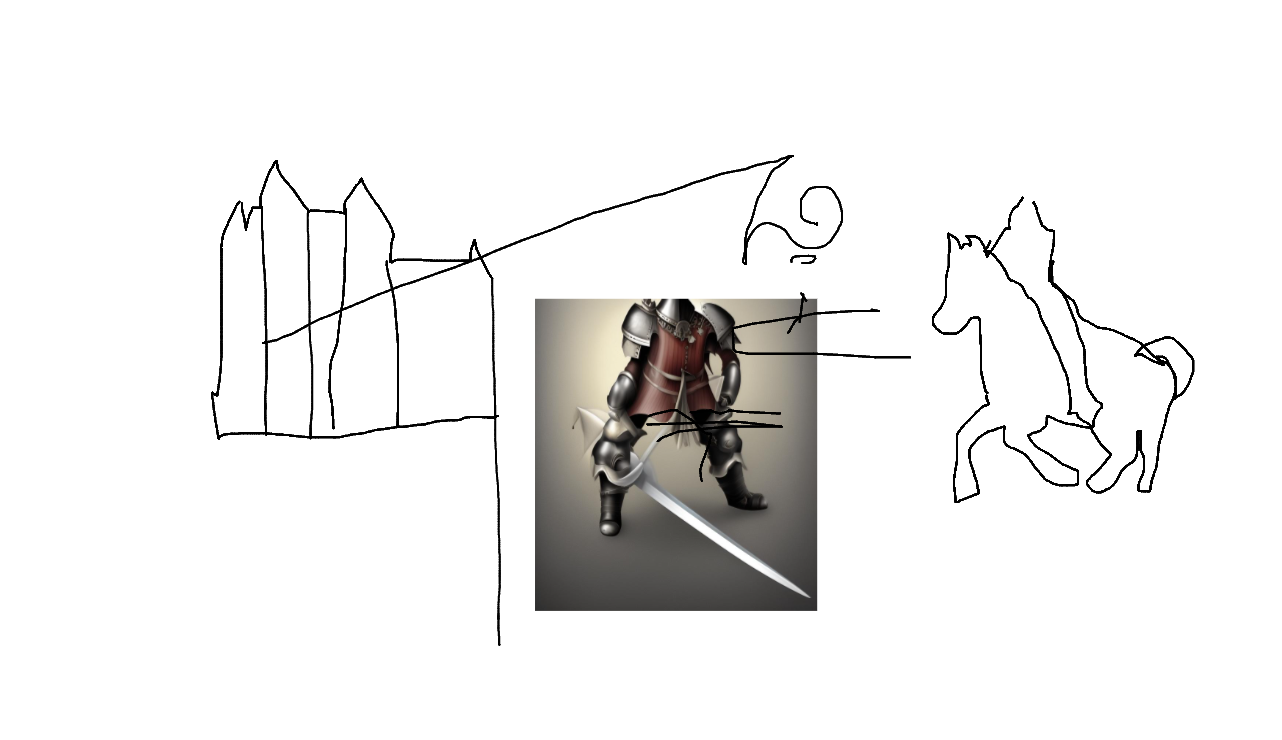} \\
    2 \includegraphics[width=0.4\linewidth]{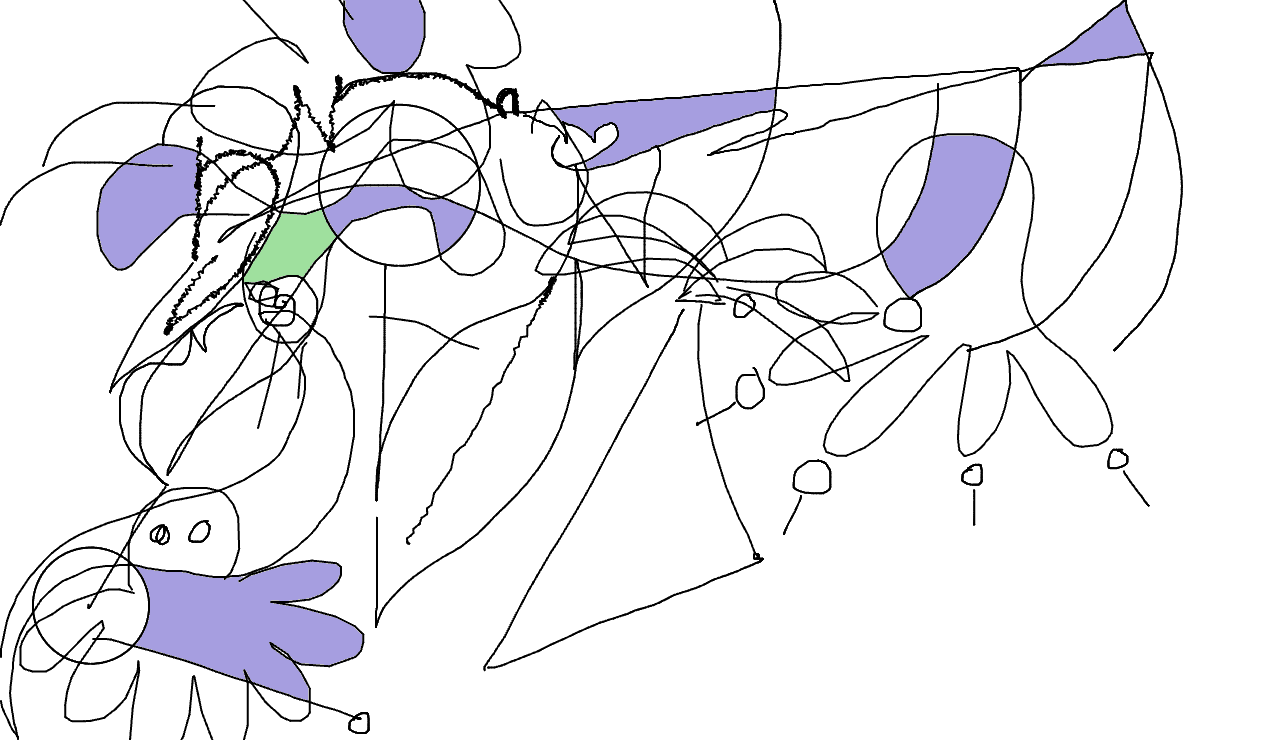} & 7 \includegraphics[width=0.4\linewidth] {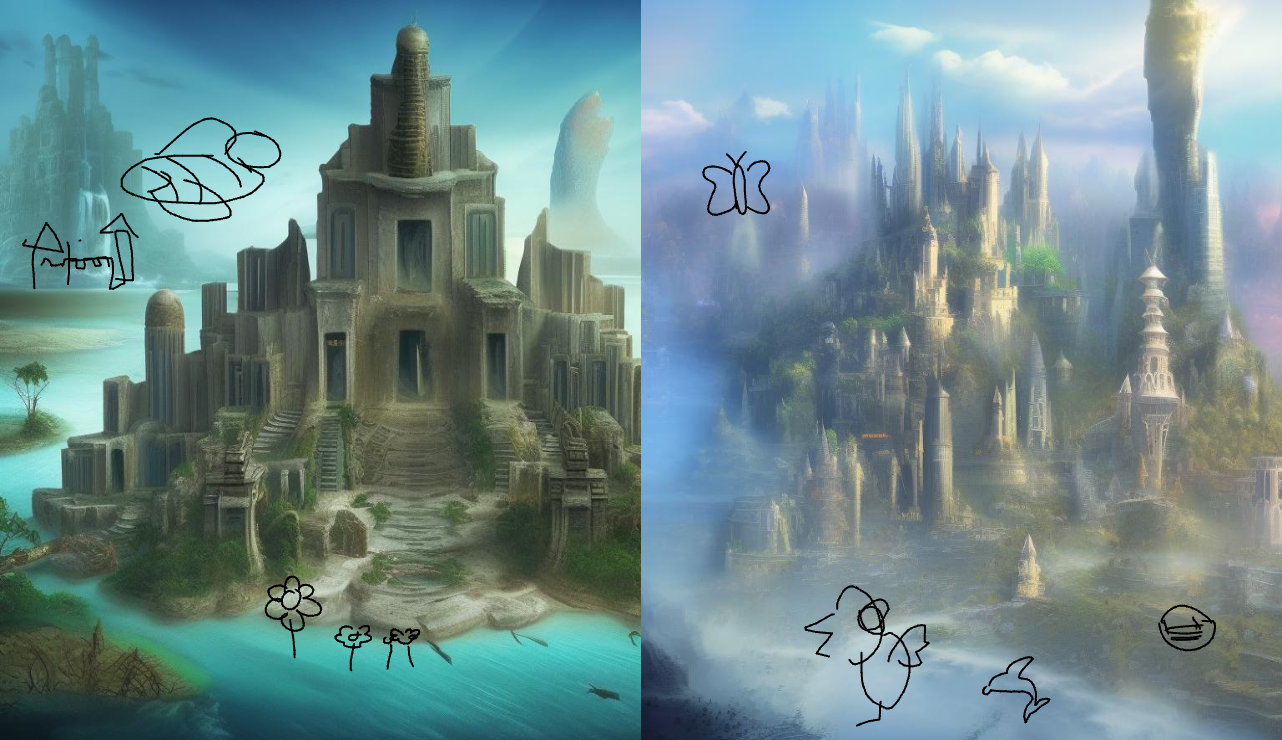} \\
    3 \includegraphics[width=0.4\linewidth]{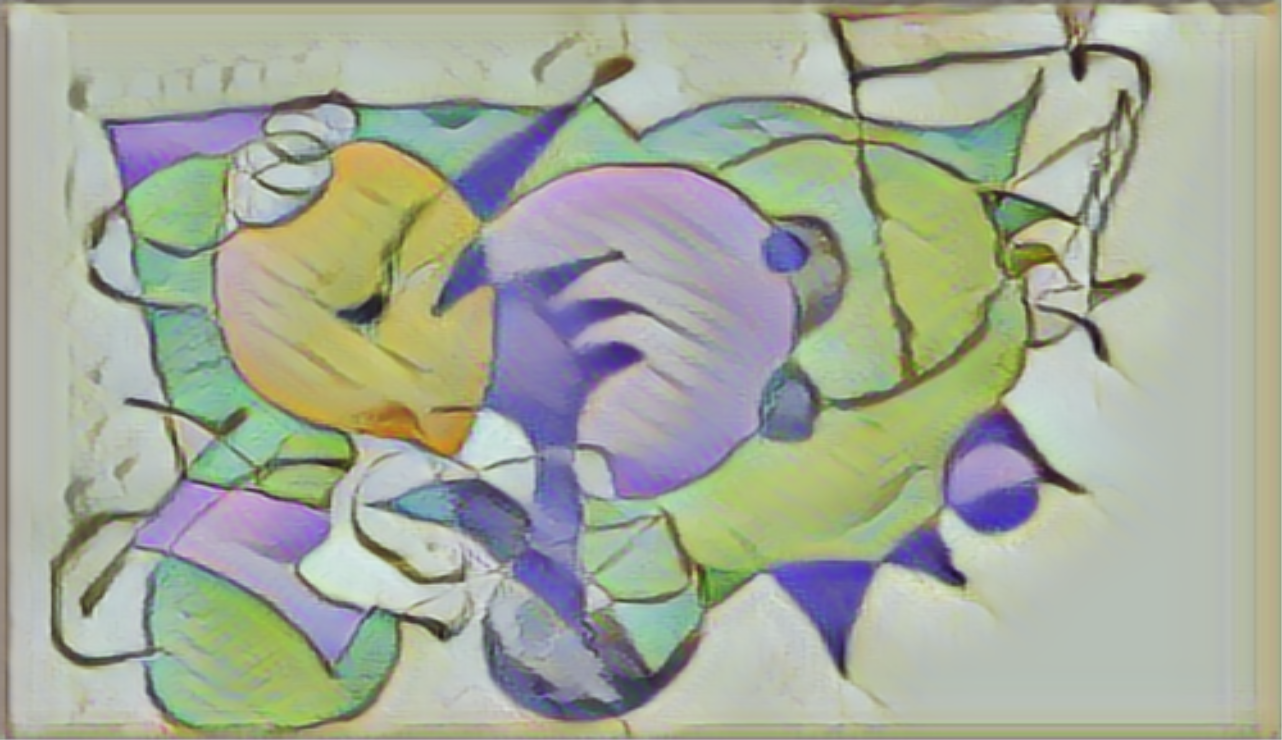} & 8 \includegraphics[width=0.4\linewidth] {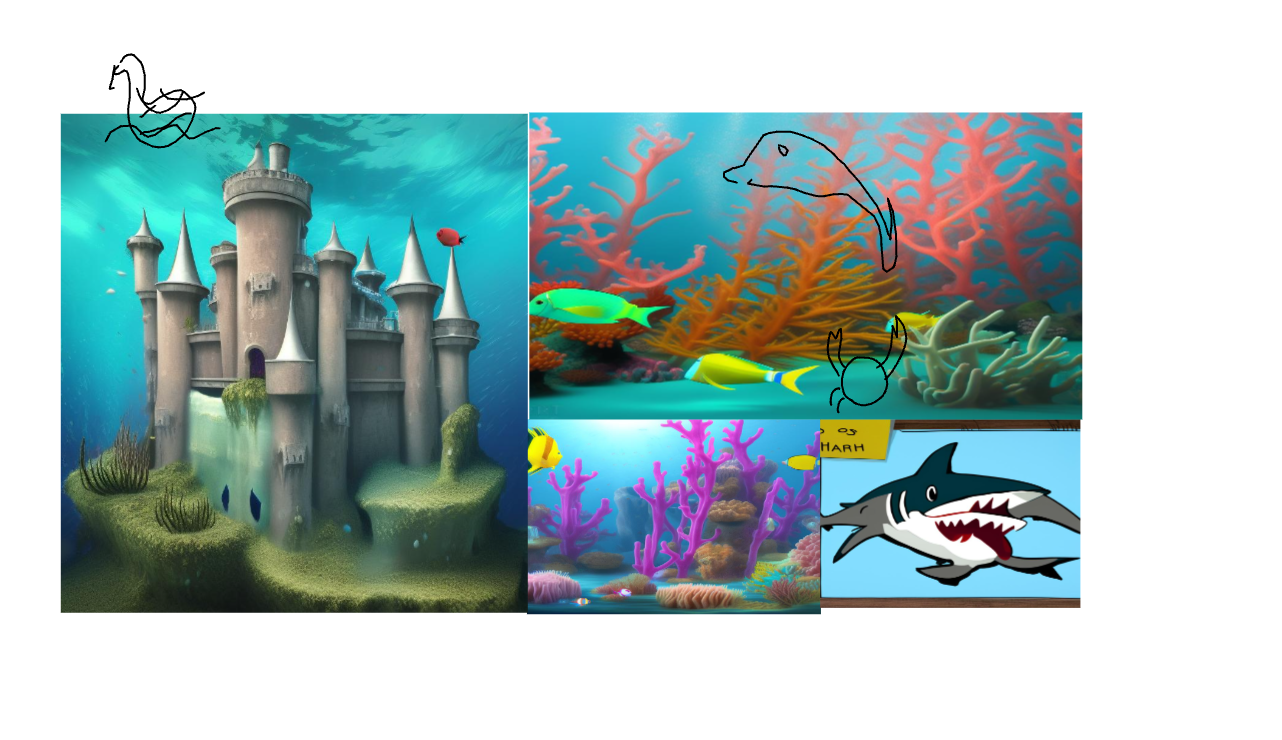} \\
    4 \includegraphics[width=0.4\linewidth]{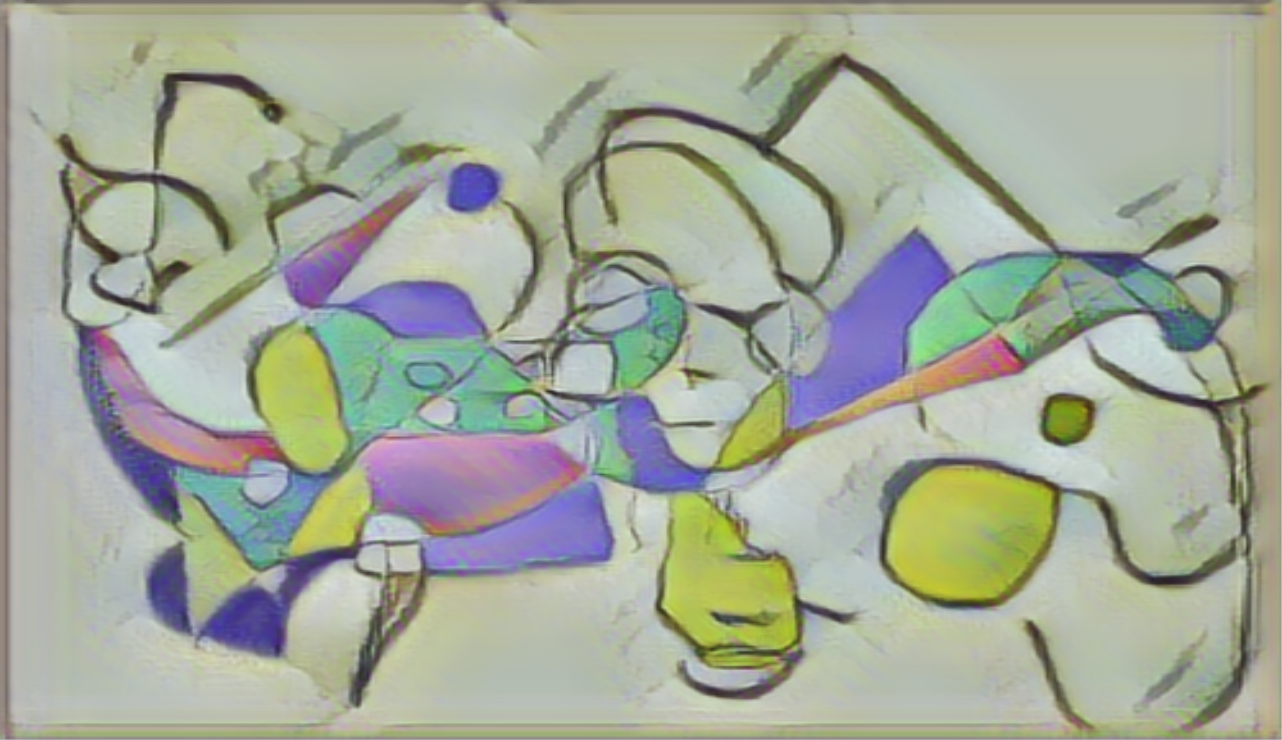} & 9  \includegraphics[width=0.4\linewidth] {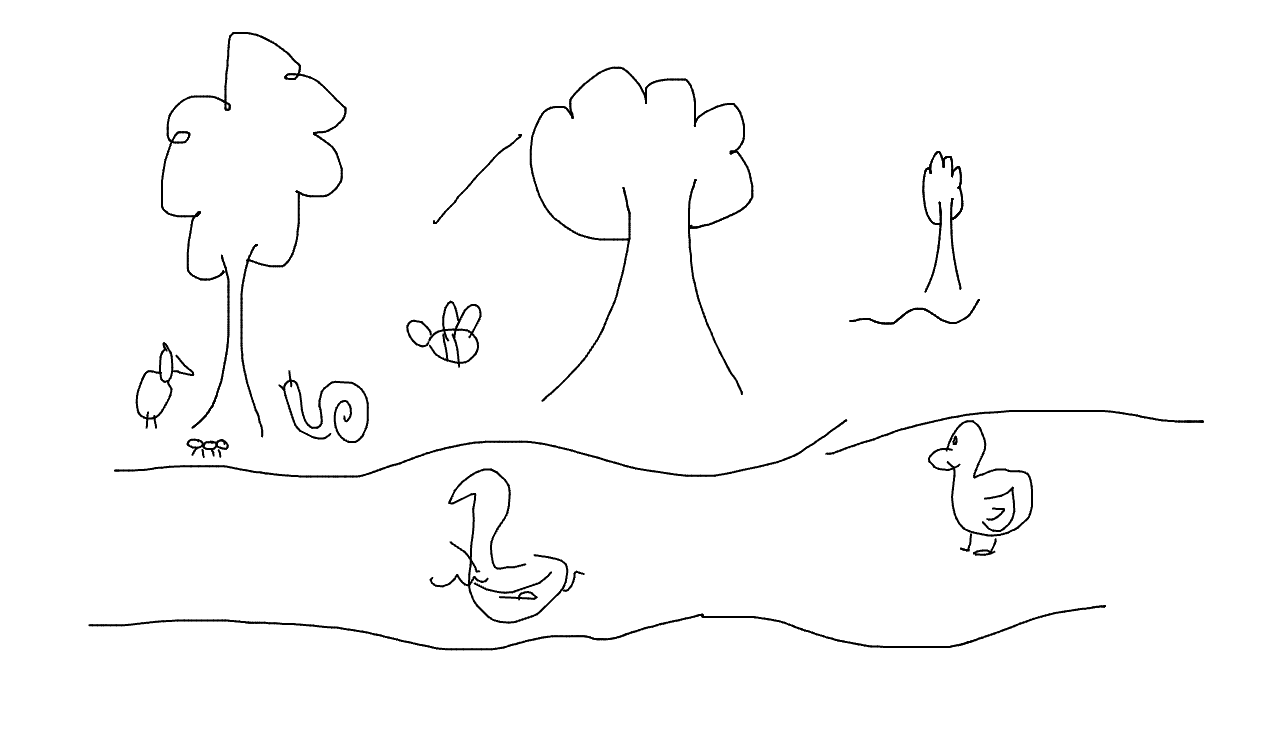} \\
    5 \includegraphics[width=0.4\linewidth]{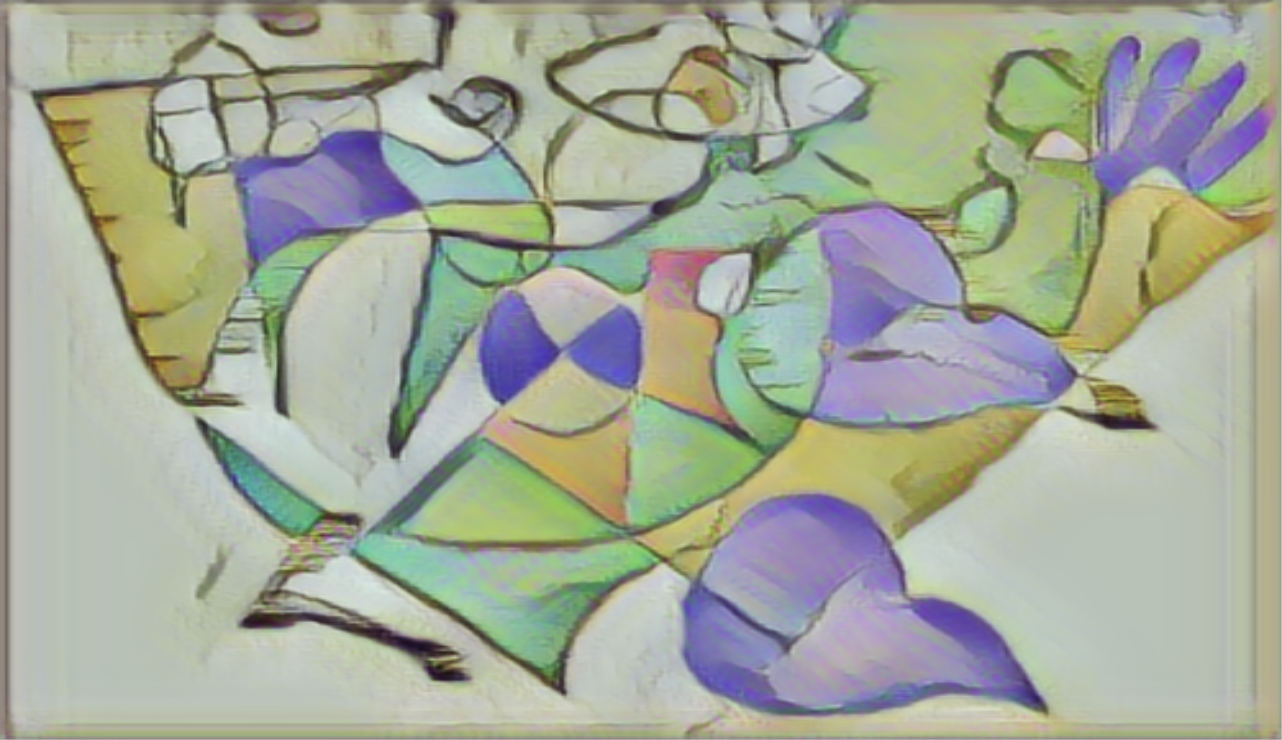} & 10 \includegraphics[width=0.4\linewidth] {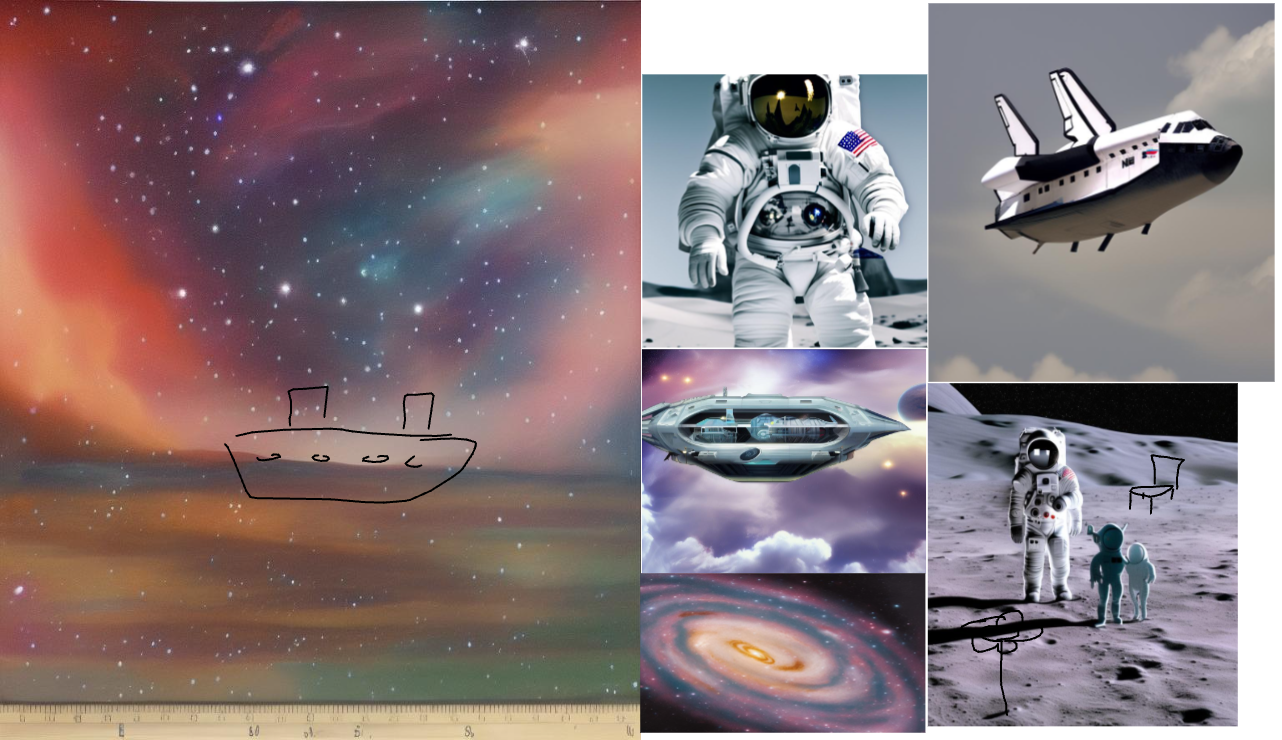} \\
    \bottomrule
\end{tabular}
\end{table}

Table \ref{tab:Artworks} shows the artworks generated during the abstract and representational co-creative sessions. Each artwork displayed is the final product of a 5-minute session. The left column is the abstract condition, and the artworks feature flowing lines. The first two artworks are drawings with some colored in regions, while the last three artworks use the stylization feature of the AI Drawing Partner to add a painterly effect to the artwork. Artworks 3-5 use the Kandinsky style, so they bear some resemblance to each other. The representational artworks use the sketch and image generation capabilities of the system. The first artwork (number 6) features tracing, while 7, 8, and 10 feature images generated by the system. Artwork 9 depicts a river scene. 

\section{Results}
The AI Drawing Partner is a quantified co-creative system that records, models, and visualizes the interaction dynamics happening on the platform. There is a companion app to the AI Drawing Partner that takes the data generated by the AI Drawing Partner and analyzes it to find significant differences between two groups. The output from the companion app is shown in the Appendix and described below. We now discuss differences in wait code, execute code, interface manipulation code, and communication code as described in Table \ref{tab:interactionMode}. 

The user’s execute code (e.g. drawing action) is significantly different between the two groups (p=.039). This difference is to be expected because in the abstract condition, the user drew more, whereas in the representational drawings, the user had the system generate images and sketches. The interactional unclamp (waiting) is not significantly different between the groups. Communication is significantly different between the two conditions (p=.003), with the representational group having a mean of 170 counts of the code applied, while the abstract group has a mean of 8. During the representational sessions, the user was communicating to the agent what they wanted the agent to do, while in the abstract sessions, the user voted to communicate to the agent. The interface manipulation code is significantly different between the two groups (p=.009), with an average of 56.8 for the abstract group, and 10 for the representational group. This was probably due to the editing the brush, filling, and stylizing done in the abstract sessions. The average coded value (e.g. the average of all the codes applied) is significantly different between the two groups (p=.0003). The abstract group had a mean of -.18, and the representational group had a mean of .22, demonstrating more fluid execution of drawing in the abstract versus more regulating the interaction in the representational sessions (e.g. manipulating interface and communicating). 

The slope of the creative sense-making curve generated by the code values is significantly different between the two conditions (p=.002). The slope of the abstract group’s CSM curve  is -.19 versus .18 for the representational group. This is due to more drawing in the abstract sessions and more communicating to the agent in the representational sessions. 

A line is calculated as the content between a pen down event and when the user raises their pen. The total lines metric shows the significant difference in the amount of content produced (sketched content) with a p-value=.005. The mean total lines for abstract was 18.8 versus 4.4 for representational. The total line length for the groups reinforces this difference with a mean 7591 pixels for abstract and 2521 pixels for representational (p=.012).

The number of turns between the two groups was significantly different (p=.0001) with a mean of 14.4 for abstract and 1.4 for representational. Generating sketches and images does not count as a turn in the system, which means the actions the user was taking did not contribute toward the turn count. When the system generates requested images and sketches it does not count as a turn because the system is not taking any initiative to act, it is merely serving as a tool. Thus, requesting to sketch images and sketches are considered part of the user's turn. Since the user was mainly requesting objects via text prompts and not sketching during the sessions, the number of new ideas, accepted ideas, and elaborations is also significantly different. The abstract group had a mean of 7.6 interaction couplings versus 1 for the representational group (p=.004). The representational group had a mean of 4.2 objects requested versus 0 for the abstract group (p=.05). 

The agent’s execute code was not significantly different for the two sessions. This is interesting as it shows the agent was consistently drawing in the two conditions. The agent’s wait code is significantly different between the conditions (abstract mean: 420.2; representational mean: 250.8; p=.02), which demonstrates the agent spent more time waiting while the user was drawing in the abstract condition. The communicate code is significantly different between the two groups (abstract mean: 73.8; representational mean: 217; p=.03) as the agent was responding to user requests and describing what it is doing to the user in the representational sessions. The slope of the agent’s CSM curve is not significantly different between the two groups, which shows that it was acting consistently throughout the sessions. 

The total agent lines are not significantly different between the sessions, but the total line length is significantly different (p=.03) with a longer total length in the abstract session (22,287 pixels) versus the representational session (6,447 pixels). The average agent lines per turn is significantly different (p=.02) with a mean of 4.4 lines in the representational condition and  2.4 lines in the abstract condition. This difference could have been because the system was generally drawing objects in the representational condition, which have multiple lines. The average line length per turn for the agent is not significantly different between the two conditions.

The number of agent turns is significantly different between the sessions (p=.01) with 15.4 turns in the abstract and 6.4 turns in the representational sessions. This is consistent with the user taking fewer turns in the representational sessions. Since the number of turns is so different the number of accepted ideas and elaborations is significantly different as well. 

\subsection{Creative Sense-Making Curves}
By summing the raw code application values, it is possible to produce a curve depicting trends in the data referred to as the creative sense-making curve. When the curve is trending downward, that means the user is drawing. When the curve is trending upward, that means the collaborator is either communicating or manipulating the interface (i.e. regulating interaction in some manner). When the curve is trending horizontal, it means the collaborator is waiting.

\begin{table}
 \caption{Creative sense-making curves for the abstract (left column) and representational (right column) co-creative drawing sessions}
  \label{tab:Graphs}
  \begin{tabular}{cc}
    \toprule
    Abstract & Representational \\
    \midrule
    1 \includegraphics[width=0.4\linewidth]{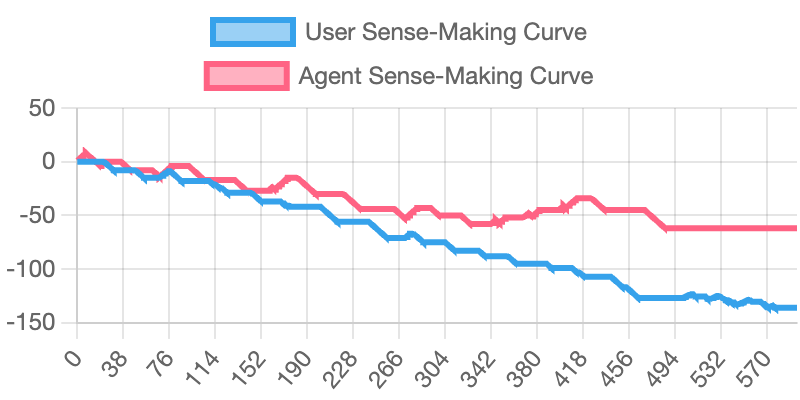} & 6 \includegraphics[width=0.4\linewidth] {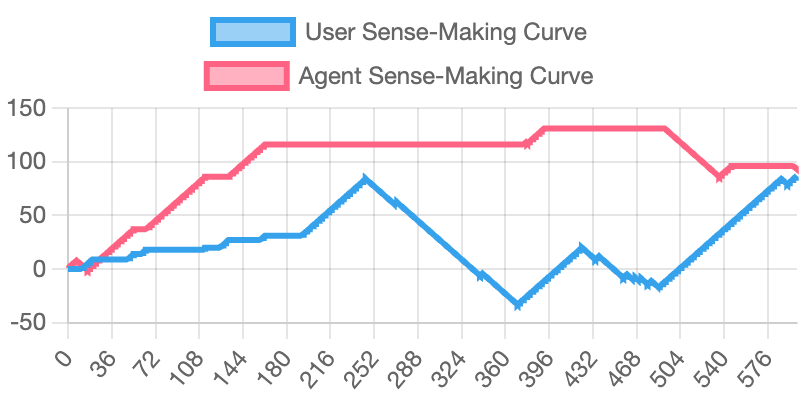} \\
    2 \includegraphics[width=0.4\linewidth]{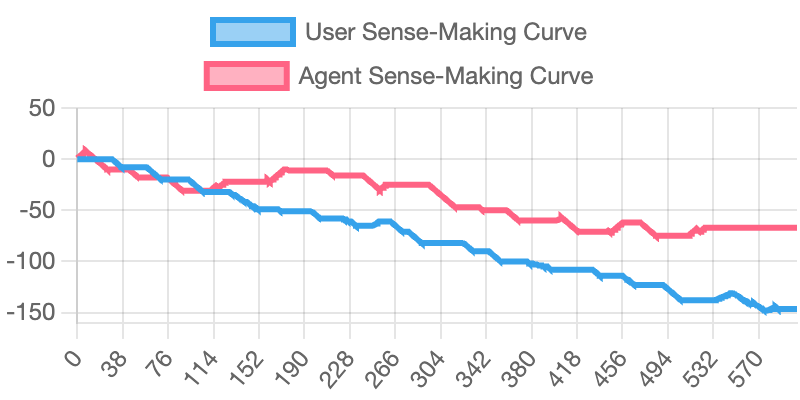} & 7 \includegraphics[width=0.4\linewidth] {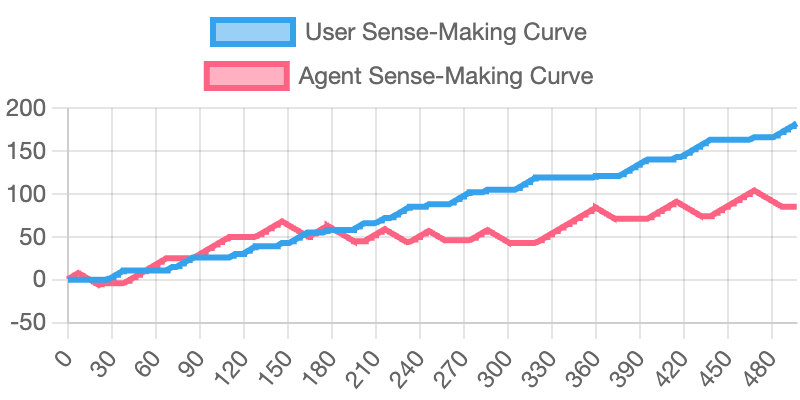} \\
    3 \includegraphics[width=0.4\linewidth]{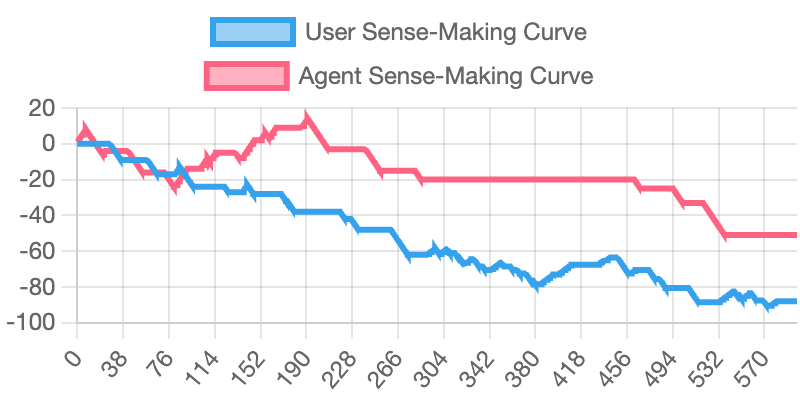} & 8 \includegraphics[width=0.4\linewidth] {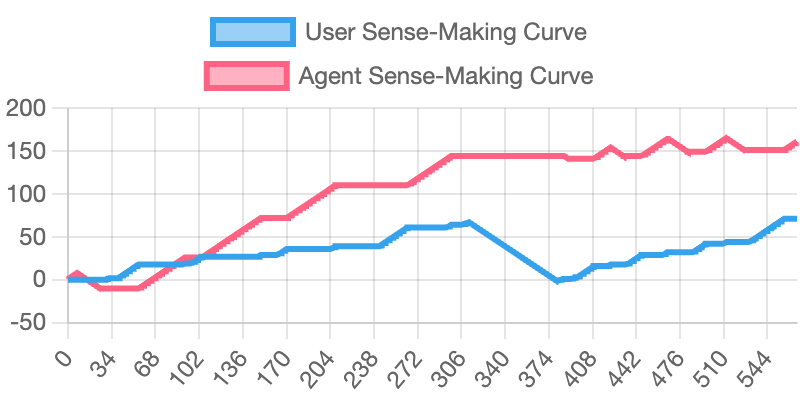} \\
    4 \includegraphics[width=0.4\linewidth]{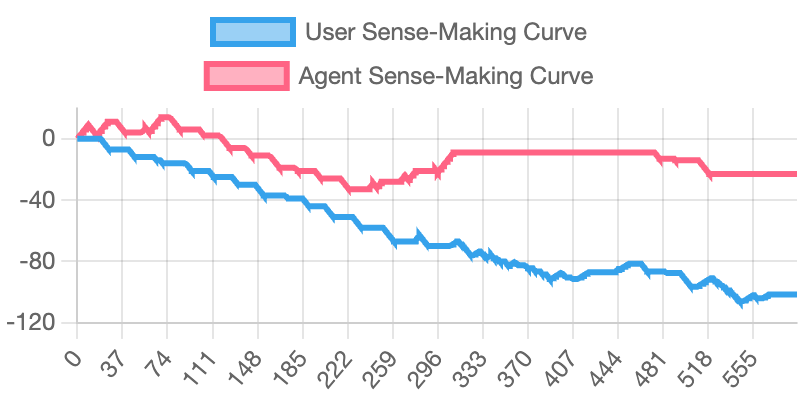}& 9  \includegraphics[width=0.4\linewidth] {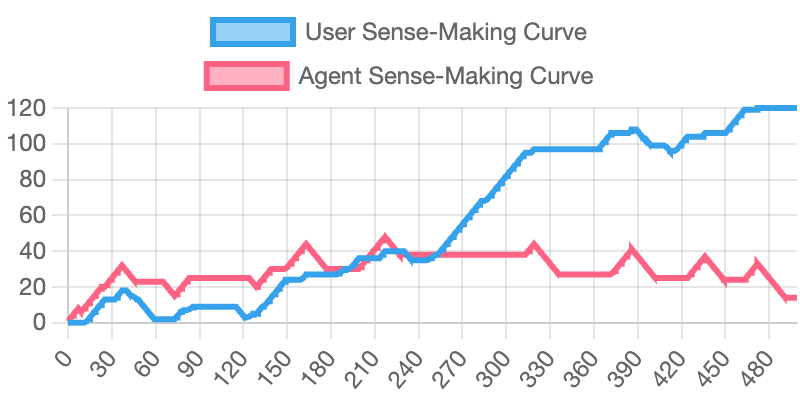}\\
    5 \includegraphics[width=0.4\linewidth]{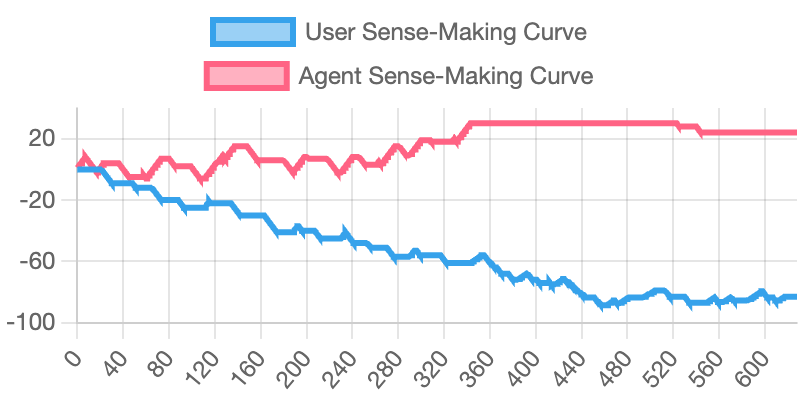}& 10 \includegraphics[width=0.4\linewidth] {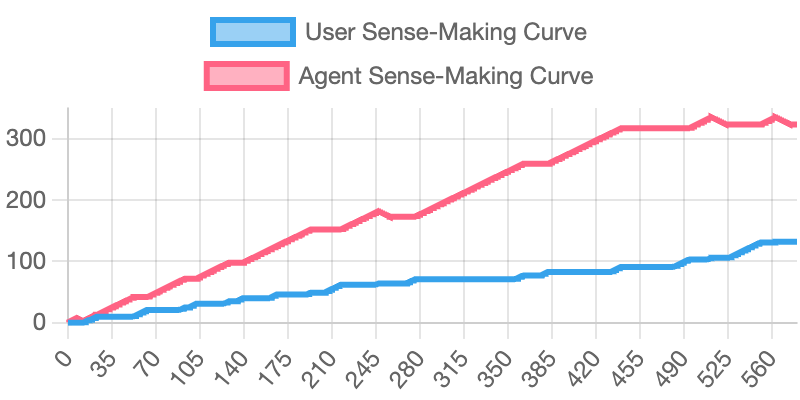}\\
    \bottomrule
\end{tabular}
\end{table}

The CSM curves for the abstract group show consistent downward trends in a stepwise fashion, which indicates turn taking. The representational group curves are trending upwards in different manners. Session 6 has two spikes, which represent requesting an image and then sketching over that image. Sessions seven and ten the user was only generating sketches and images, so the upward trend is consistent. Session eight features one long drawing action, which is the dip in  the curve between 306 and 374, while in session nine, the user taught the system how to draw a tree, which is the large upward trend at timestamp 270.

\subsection{Trend Classification}
Since the CSM curve is a continuous linear time-series dataset, it is possible to perform time-series analysis, such as stock market technical analysis, on the data to identify upward, downward, and horizontal trends in the data. This will enable the identification of sequential trends in the interaction dataset. While applying stock market analysis to creativity research is novel, the format of the data produced by the CCSM enables new types of analysis, and stock market analysis is effective at identifying trends in temporal data. It is not as strong in identifying patterns or sequences of trends in the data, which would be the next step in the analysis and could be performed by machine learning.

\begin{figure}[h]
  \centering
  \includegraphics[width=\linewidth]{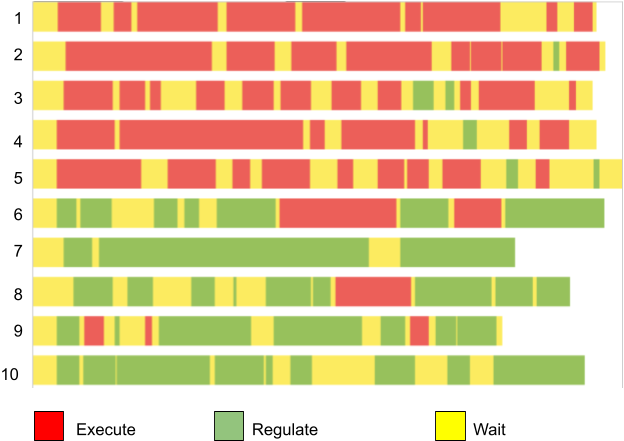}
  \caption{The trend sequences for the 10 co-creative drawing sessions. The trends were identified using stock market technical analysis.}
  \Description{}
  \label{TrendSeq}
\end{figure}

For our purposes, a buy signal would equate to regulating the interaction (manipulate the interface or communicate), a sell signal would equate to executing drawing actions, and a hold signal would correspond to waiting. The analysis uses the Moving Average Convergence Divergence (MACD) technical analysis indicator and the Exponential Moving Average (EMA) to determine buy, hold, and sell signals. In this analysis, a period is a step in the time interval in the data, which would represent 0.5 seconds. The 12-period EMA (e.g. 12 temporal intervals) is calculated and subtracted from the 26-period EMA to yield an MACD line. Then, a signal line is produced, which is the 9-period EMA of the MACD line. When the MACD line crosses above the signal line, it is a buy indicator, while when it crosses below the signal line, it is a sell indicator. The AI Drawing Partner companion app calculates these indicators and visualizes them, as shown in Figure \ref{TrendSeq}.

In Figure \ref{TrendSeq}, a trend classification is made for every time step in each data set. Then, trends that are the same in a sequence are grouped. The length of each grouping determines the width of the colored rectangle. Fluidly executing drawing actions is red. Regulating the experience by manipulating the interface or communicating is green. And whenever the user or agent is waiting it is yellow. The trends for the abstract group (sessions 1-5) are mostly execution-waiting cycles (interactional coordination), while the trends for the representational group (sessions 6-10) are mostly regulate-wait cycles (functional coordination). 

Visualizing interaction trends in this manner enables comparative analysis within and between participants. The analysis can examine the co-creative experience in phases, with the trend sequence being cut into quarters. Then, the frequency of trends can be noted for the different phases of co-creation. It is also possible to visually identify which trends are more frequent between participants, such as between group 1 and group 2.

The AI Drawing Partner companion app automatically generates this visualization on CSM curves inputted into the system. It performs the MACD analysis on the curves and visualizes the results, which can then be labeled in another software application, such as Photoshop or Google Drawing.

\section{Discussion}
The AI Drawing Partner quantifies and models the co-creative process using the co-creative sense-making framework (CCSM). The data it generates enables new forms of analysis of co-creation, such as visualizing interaction dynamics (CSM curves in Table \ref{tab:Graphs}) and identifying interaction trends (Figure \ref{TrendSeq}). The significant differences between the two groups (representational and abstract) correspond to qualitative differences in the co-creative experiences of the user, e.g. the user was interacting qualitatively differently in the two conditions and the metrics were able to capture this difference. 

\subsection{Human-AI and AI-Human Communication}
The significantly different communication coding values between the two conditions (p=.003) can be examined more closely. In the abstract condition, there was minimal communication (group mean of 8 counts of the code applied). The only time spent communicating was voting. In the representational condition, however, the user was requesting sketches and images more and communicating to the agent with a group mean of 170 counts. The COFI framework suggests that increased levels of communication and means of communication can lead to increased engagement \cite{rezwana2022designing}. There is a design opportunity here to increase modalities of communication in more abstract conditions. One avenue to explore this is integrating a large language model (LLM), such as ChatGPT, into the system to allow the user to discuss and reason about the artwork with the system. When using an LLM chat interface the user can tell the system what they are drawing, for example, and it can learn that object, pattern, or structure and store it in the database for future use. 

\subsection{Wait time and Dyadic Creativity}

In this study, "wait time" was coded as any pause, hesitation or non-action the mode of cognition is considered to be a interactional unclamped and the agent is actively engaged in sense-making. However, it may prove meaningful to make a distinction between unintentional wait time for the algorithm to process (e.g. wait time for image generation) and when the user intentionally pauses to reflect on the creative process while still in their own turn. 

In dyadic, human-human creativity, waiting for a partner to act introduces a pause that can be filled with anticipation, offering a moment to reflect on the evolving collaborative process. This waiting period can foster a deeper understanding of one's partner's thought process and approach to the task, enriching the collaboration. It can also be a time for individual ideation, where the "silent gaps" become a time for an individual to adapt in response to anticipated contributions. As research indicates that dyadic creativity can attain a higher level of originality and experience stronger feelings of stimulation, enjoyment, and originality of expression \cite{torrance1971stimulation, rosenberg2022social}, working towards a system which mimics dyadic creativity may be advantageous. Thus, there is a design opportunity here to investigate and optimize the wait response time from the system and the interaction narrative to support the dynamics of give-and-take during these pauses which may also strengthen the sense of partnership between end user and agent.

\begin{figure}
    \centering
    \includegraphics[width=.5\linewidth]{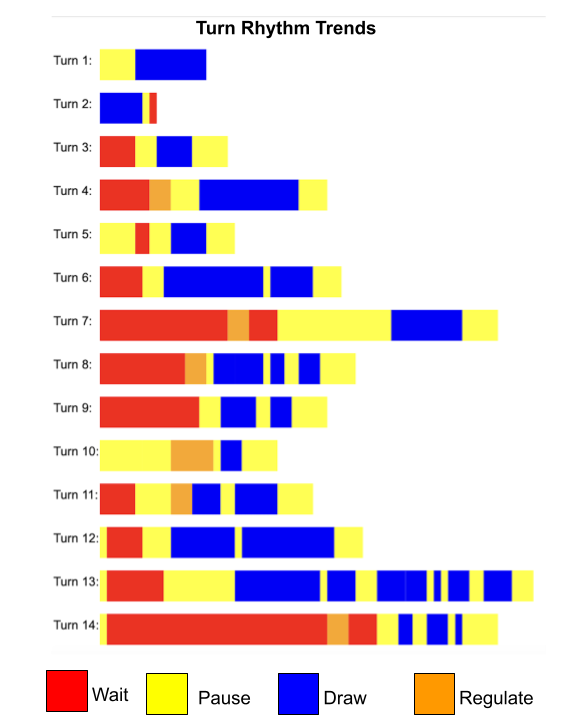}
    \caption{Visualization of the rhythm of turns in a 5 minute co-creative drawing session.}
    \label{fig:turnRhythm}
\end{figure}

As a proof of concept, the AI Drawing Partner was updated to reflect this distinction and another five minute drawing session was conducted to yield the turn rhythm trends in Figure \ref{fig:turnRhythm}. When the user is waiting, the agent is drawing and the user is not performing any actions. Pausing is defined as any time of non-action when the agent is not drawing. Drawing is defined as any time the pen is touching the digital canvas. Regulate is defined as interface manipulations (e.g. opening the settings menu) and communicating with the agent (e.g. voting, requesting sketches/images). In Figure \ref{fig:turnRhythm}, the length of each turn is visualized, as well as what composes each turn. The agent typically begins each turn. The amount of time the agent spends drawing is typically approximately equal to the length of the user's last turn, such as in turn 4, 6, 7, 8, 9, 11, 12, and 14. There is regulation directly following the agent's turn in turn 4, 7, 8, and 14, which indicates the user's voting behavior. 

This type of temporally sensitive visualization of the co-creative process would not be possible if the AI Drawing Partner was not a quantified co-creative system. However, since the system is continuously collecting data about the interactions of the user and agent, this data can be utilized in time sequence analysis of the co-creative process. 

\subsection{Broader Impact on Human-AI Systems}
The development of co-creative interfaces, such as the AI Drawing Partner, address several significant challenges inherent in the fields of human-Centered AI and hybrid intelligence \cite{rafner2023creativity} while also grappling with the ethical dilemmas these fields present. In particular, the AI Drawing Partner's interface aims to create simultaneously high degrees of automation and human control as well mutual learning in which both human and machine processing and outputs adapt to optimally empower each other. 

With respect to the simultaneously high degrees of automation and human control, the AI Drawing Partner enables text-to-image or voice-to-image generation in the context of creating a shared artwork. This means the images can be directly manipulated and edited, instead of the static images that are typically generated via text prompting. This text or voice prompt enables high levels of automation, and the instructions provided to the agent about placement and size of the generated image enable a high degree of control. The system also features different drawing modes that serve to constrain the generative space. These modes provide the user with control, and help them develop a mental model of how the system behaves in each mode and overall. 

In the AI Drawing Partner there are several ways in which the human user and the AI agent learn and adapt from each others input. The user's positive and negative feedback (up/down vote) provide a means of interactive machine learning \cite{fails2003interactive, dudley2018review}. The agent's actions that result from the user's feedback form another feedback loop for the human that informs their sense-making and learning processes. Additionally, there are mechanisms of instruction for both the user and agent. The user instructs the system to generate creative output, and the system seeks instruction from the user about how to place and size that creative output. Furthermore, integrating a LLM into the system, as mentioned above, would allow also support the system learning personalized, specific, objects, patterns, or structures and store them in the database for future use. 

Importantly, because the AI Drawing Partner is explicitly designed as a research platform, it allows other researchers in various domains to design experiments to further explore these important elements of human control, increased automation and mutual learning without needing to develop their own systems. The benefits of a standard sense-making framework and data collection technique include enabling researchers to compare results within creative domains and across creative domains. Standardizing research methodologies will help the young field of co-creative AI be more methodical. The use of the AI toolbox approach enables the addition of new AI models as they become available. Instead of creating entirely new systems based on advances in AI, these models can be directly added to the system architecture by placing them in the AI toolbox. Finally, the automatic data collection and visualization of the AI Drawing Partner significantly reduces the time it takes to analyze co-creation. Automatic analysis reduces human error and biases in the coding process. It also eliminates the need for inter-rater reliability, as the system is quantifying interaction dynamics as they occur in the system. 

Increasing scrutiny is being directed toward the ethical challenges posed by human-AI technologies and co-creative methods are no exception \cite{10.1145/3544549.3585657}. Firstly, there is an on going debate about the issues of AI systems poised to simply automate tasks, potentially deskilling workers \cite{rafner2022deskilling} and ultimately replace them all together. The AI Drawing Partner, developed as a real-time co-creative system under the principles of human-centered AI and Hybrid Intelligence seeks to augment human creativity rather than replace it. Additionally, genAI models are trained on publicly accessible creative works. The creators of these models often fail to adequately credit the creators of the data they use, leading to potential intellectual property violations and broader issues of wage theft. Additionally, by disproportionately representing dominant styles, norms and content in their training data these technologies tend to reinforce existing biases, further marginalizing underrepresented groups \cite{lawton2023tool}. With respect to our system, the CNN sketch recognition module and Sketch-RNN sketch generation model were trained on the Google Quick Draw dataset \cite{quickdraw}, which contains 50 million drawings and has 345 object categories. Over 15 million players have contributed to the Quick Draw dataset in numerous countries. Each drawing has metadata describing what the user was asked to draw, and their country of origin. Each category contains a variety of sketch representations of the category from different cultures around the world. The Quick Draw dataset is publicly available \cite{quickdraw}, and serves to democratize the training of sketch-based machine learning and artistic projects. However we acknowledge that the text-to-image feature of the AI Drawing Partner utilizes Stable Diffusion, meaning some of the imagery is inspired by uncredited contributions.

\section{Future Work}
Here below we outline directions of future work categorized by analysis methods, technical development and experimental set-ups. 

\subsection{Analysis methods}
In addition to stock market technical analysis, it is possible to perform a machine learning analysis of the CSM curve to identify trends and patterns. A retrospective protocol analysis can be performed with the user to determine what their intention was throughout the co-creative process (e.g. abstract drawing, drawing an object, drawing a scene, filling objects). The intention would serve as the label for the CSM curve, such that the algorithm would learn the associations between trends in the CSM curve and the user’s intention. 

The interaction model produced by the AI Drawing Partner can also be utilized by the system to influence its creative behaviors. It can use machine learning to automatically recognize trends and adapt its behavior to align with the current trend. For example, ideation might have a different interaction pattern than refinement. If the system detects that the user is currently ideating, it can generate more diverse responses. If the system detects the intention of refinement, then it can converge on a more limited subset of actions that align with what the user is refining. 

\begin{table}
 \caption{Interaction coding scheme for co-creative programming.}
  \label{tab:Coding}
  \begin{tabular}{clc}
    \toprule
    Code & Value & Behavioral Marker \\
    \midrule
    Communicate & 1 & Chatting with the co-pilot \\
    Manipulate Interface & 0.5 & Scrolling, searching, copy/paste \\
    Wait & 0 & Pausing, hesitating, non-action \\
    Execute & -1 & Fluidly typing code \\
    \bottomrule
\end{tabular}
\end{table}

The CCSM is a domain independent cognitive framework that can be adapted to other co-creative domains \cite{davis2024creative}. For example, it could be applied to co-creative programming. The interaction coding scheme would be as shown in  Table \ref{tab:Coding}. The coding procedure can happen automatically (by the system) or be applied manually through video coding of the interaction. 

Quantified co-creative systems introduce an experiment and analysis pipeline that significantly shortens the time it takes to conduct and analyze a study. The system automatically quantifies, models, and visualizes the co-creation data. The companion app analyzes all the generated data and finds statistical significance as well as identifies trends in the data. 

\subsection{Technical development}
The technical future direction of the AI Drawing Partner is to design the system to use the creative sense-making curve as a model of user behavior and respond according to this model. The system can recognize different creative intents of the user (e.g. ideating, refining, coloring) based on the shape of the creative sense-making curve. Once the system detects the user’s creative intent, it could align its intention to the user’s, promoting participatory sense-making. The project will also explore situations in which it is appropriate to diverge from the user’s intention and introduce a new creative direction in the artwork. These types of antagonistic contributions have the capacity to catalyze the creative process because the user has to make sense of them and integrate them into their artwork \cite{davis2011computing}. However, if there are too many antagonistic contributions, the user may become frustrated and think the system is behaving erratically.

\subsection{Experimental set-ups}

Future work will evaluate the AI Drawing Partner with two user groups: experts and novices. The participants will first be oriented to the tool with an introductory video explaining the functionality. Each group will then interact with the tool for 15 minutes to create any artwork they choose. Following the drawing session will be a retrospective protocol analysis where the participant and researcher review a video of the interaction and the participant explains their cognitive processes during the co-creative session. The user will then fill out a survey focusing on the degree of creativity support the tool offered and elements of human-AI collaboration. This will be followed by a semi-structured interview focused on the sense-making and participatory sense-making processes the user engaged in, particularly what enabled participatory sense-making.  

Future work will also explore the application of the AI Drawing Partner in educational contexts. This work will evaluate whether the tool can teach creativity, social collaboration skills, and computing concepts in a fun and engaging manner. The target audience for this evaluation will be students aged 7-12. The students will also watch an introductory video to familiarize them with the functionality of the system. Next, the students will engage in a co-creative drawing with the tool for 15 minutes. This will be followed by a retrospective protocol analysis, a survey, and an interview. The participants will be split into two age groups: 7-9 and 10-12, to compare the interaction dynamics of these different age groups. The analysis will evaluate the creative outcomes, the social interactions between the student and the co-creative agent, and the student’s comprehension of how the AI system works. 

Other researchers in the field can conduct studies with the AI Drawing Partner system, which makes the field of co-creative AI more accessible to non-AI researchers. For example, psychologists could use the system to identify how interaction dynamics and sense-making patterns differ based on user demographics, such as age, gender, and occupation. Cognitive scientists could use the system to examine how users with different types of cognition engage with the system. 

The AI Drawing Partner can also be applied to mental health contexts. Art therapy has gained traction as a means of treating a wide variety of mental health problems. In a review, \cite{slayton2010outcome} found that “art therapy as a treatment modality has been isolated, measured, and shown to be statistically significant in improving a variety of symptoms for a variety of people with different ages” (p.115). Measuring the efficacy of art therapy \cite{slayton2010outcome} is critically important in demonstrating its value. There are several art-based assessment techniques, such as the Diagnostic Drawing Series (DDS) and the Person Picking an Apple from a Tree (PPAT) assessment technique \cite{betts2006art}. The formal elements of art therapy scale (FEATS) was developed as a systematic measure of  art therapy outcomes \cite{gantt1998formal}. It features Likert-scale questions that the rater fills out to analyze the drawing. The AI Drawing Partner could provide a platform with which art therapists can quantify their client’s creative processes through time. In addition to qualitative assessments, the AI Drawing Partner could provide quantified data about the drawing behaviors and interaction dynamics of the participant. The system could be used with the AI turned off to perform standardized assessment techniques, such as the DDS and PPAT, in a quantified drawing environment, or the AI could be turned on for therapeutic value to help inspire and creatively engage the client in ways that traditional art-making does not. The turn-taking interaction dynamic of co-creative AI has been found to positively engage users and help inspire them about what to draw next \cite{davis2016co}.

\section{Limitations}
A case study provides a limited capacity to evaluate the CCSM fully. A case study was selected for demonstration purposes to exemplify the CCSM analytical technique. A case study enables the analysis of individual differences throughout multiple sessions. However, a user study with expert artists would greatly enhance the validation of the CCSM. A user study would enable the aggregate comparison of different users as they engage with the system. 

The CCSM only provides quantified process data about co-creative experiences. To gain a more holistic view of co-creation, qualitative data is required. The creativity support index (CSI) can be used to determine the level of creativity support the tool offers. The CSI has six dimensions: collaboration, enjoyment, exploration, expressiveness, immersion, and whether the results were worth the effort. The CSI has been expanded into the mixed-initiative creativity support index (MICSI) that considers additional elements of human-AI collaboration, such as: communication, alignment, agency, and partnership. One can pair the CCSM analysis with the mixed-initiative creativity support index (MICSI) survey \cite{lawton2023drawing}, retrospective analysis, and semi-structured interview to gain a more holistic view of co-creation. 

The case study was for demonstration purposes only, and the data it generated may vary significantly from real world user studies, e.g. those performed in a laboratory or through online software. The user was also the designer of the AI Drawing Partner, and he was thoroughly familiar with the interface and interaction design of the system. Real world studies would feature more sense-making processes between the user and system as the user tries to figure out how the system works. 

\section{Conclusions}
This paper presented a new technique for quantifying, modeling, and visualizing the interaction and collaboration dynamics of co-creative AI. The co-creative sense-making (CCSM) cognitive framework was used as the basis for this quantification. The AI Drawing Partner implements the CCSM framework within its system design. It is a quantified co-creative system that models the co-creation happening on the platform. A demonstrative analysis was presented of 10 co-creative drawing sessions each lasting 5 minutes. There were two groups of sessions: abstract and representational. Significant interaction differences were found between the sessions that matched the qualitative differences in creative strategy between the two groups. The analysis serves to show how the CCSM can be used in co-creative systems to automatically code and analyze co-creation. Quantified co-creative systems introduce an experiment-analysis pipeline that significantly shortens the time it takes to conduct and analyze a study. The CCSM also systematizes the data collection and analysis process between different systems, especially systems in different creative domains. 

The AI Drawing Partner is a freely available and publicly accessible co-creative drawing agent and research platform that enables dynamic interaction between an AI agent and the user while automatically modeling the co-creative process. There are four main categories of data in the CCSM framework the system uses to collect data: interaction dynamics, cognitive dynamics, collaboration dynamics, and domain behaviors. Using these metrics, the co-creative process can be analyzed in a cognitively rooted manner. The framework is a domain-independent method for quantifying co-creation in co-creative AI systems. Unlike other co-creative systems, the AI Drawing Partner collects, quantifies, and visualizes the interaction dynamics of the user and agent in the co-creative session. A case study was presented of an expert artist utilizing the system to illustrate the data the system records. Future work can explore how to utilize the automatically detected interaction models to inform the creative behavior of co-creative agents to better align with the cognitive and social processes present in co-creation. 

\section{Acknowledgments}
The authors would like to acknowledge Dr. Mary Lou Maher for her help conceptualizing the paper and her contributions to the AI Drawing Partner.
\bibliographystyle{ACM-Reference-Format}
\bibliography{manuscript}

\appendix

\begin{figure}[h]
  \centering
  \includegraphics[width=\linewidth]{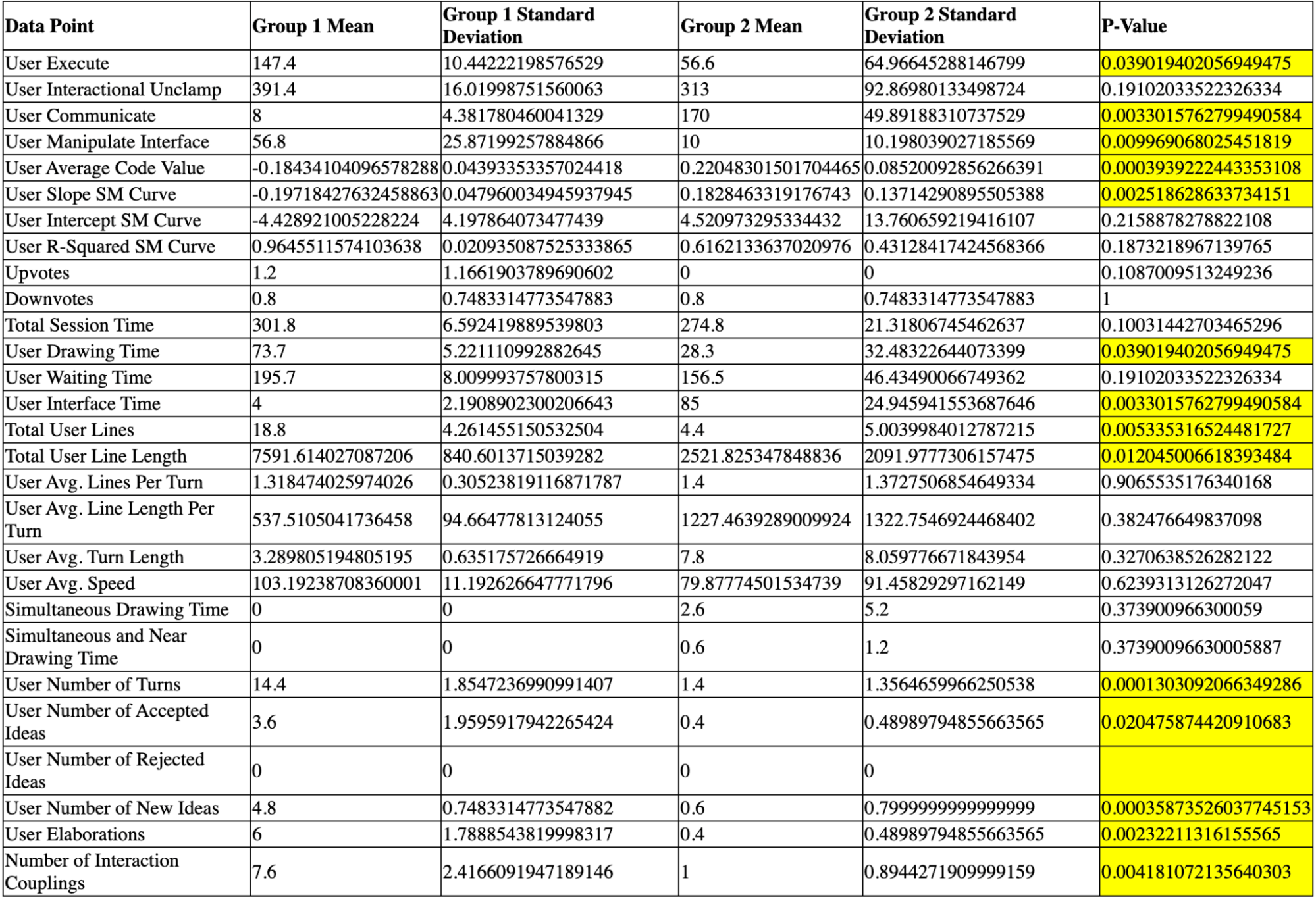}
  \caption{AI Drawing Partner companion app analysis part one. Significant differences are highlighted in yellow.}
  \Description{}
  \label{appendix1}
\end{figure}

\begin{figure}[h]
  \centering
  \includegraphics[width=\linewidth]{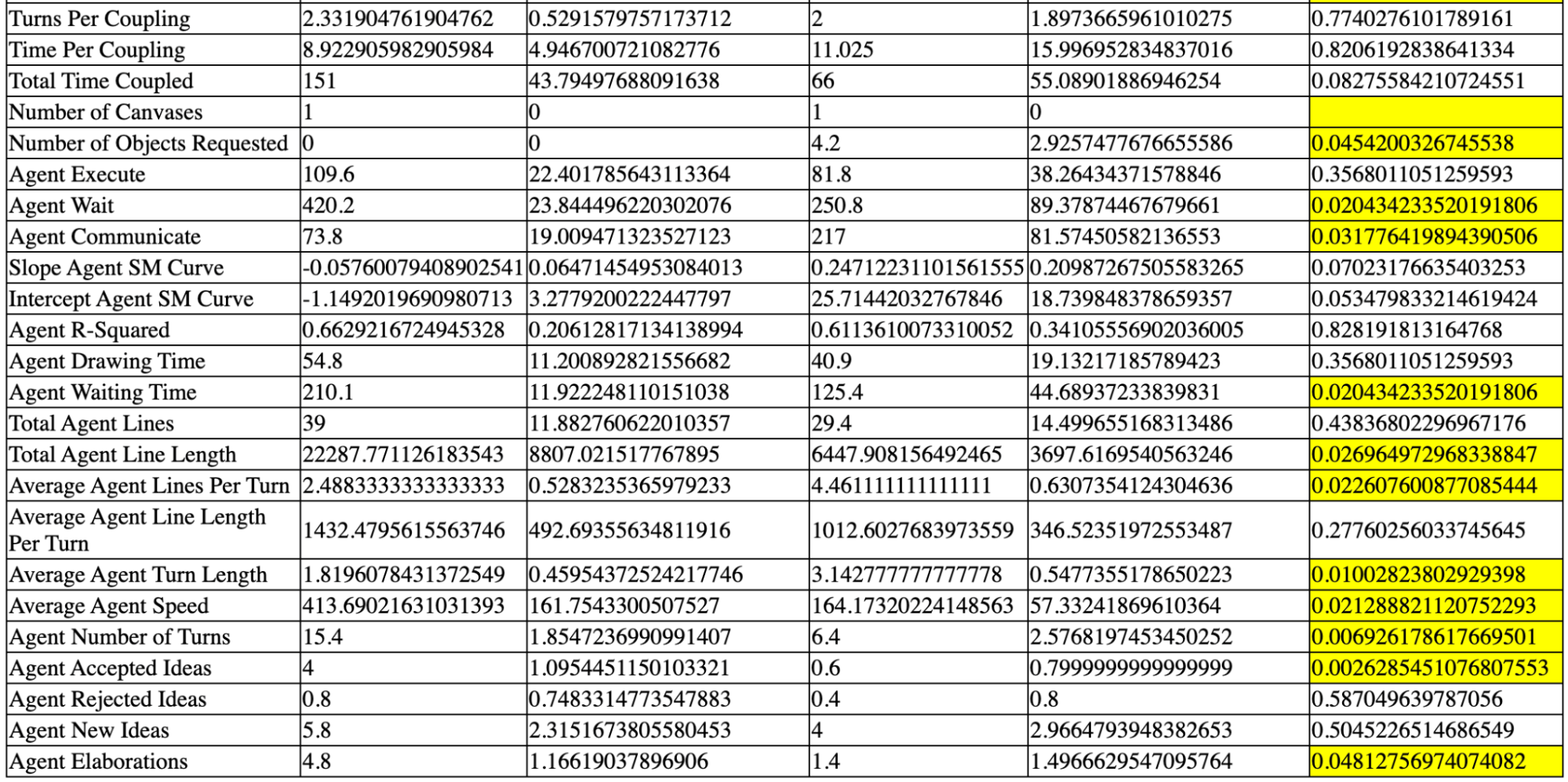}
  \caption{AI Drawing Partner companion app analysis part two. Significant differences are highlighted in yellow.}
  \Description{}
  \label{appendix1}
\end{figure}


\end{document}